\documentclass[a4paper,12pt,reqno,superscriptaddress,nofootinbib]{revtex4}
\usepackage[a4paper, left=2cm, right=2cm, top=2.2cm, bottom=2.2cm ]{geometry}
\usepackage[centertags]{amsmath}
\usepackage{amsfonts}
\usepackage{amssymb}
\usepackage{amsthm}
\usepackage{newlfont}
\usepackage{stmaryrd}
\usepackage{mathrsfs}
\usepackage{euscript}
\usepackage{siunitx}
\usepackage{comment}
\usepackage{algorithm2e}
\usepackage{graphicx,subcaption}
\usepackage{enumitem}
\usepackage{natbib} 
\usepackage{braket}
\usepackage{bbm}
\usepackage{graphicx}
\usepackage{xcolor}
\usepackage{floatrow}
\usepackage{caption}
\usepackage{hyperref}
\usepackage{subcaption}
\usepackage{hhline}
\usepackage{hyperref}
\usepackage{cleveref}
\usepackage{amsmath}
\usepackage[bb=dsserif]{mathalpha}
\usepackage{bm}

\usepackage{graphicx}   
\usepackage{amsmath,amssymb}

\usepackage{float}

\usepackage{tikz}
\usepackage{pgfplots}
\pgfplotsset{compat=1.18}
\usepgfplotslibrary{fillbetween}
\usepgfplotslibrary{patchplots} 

\usetikzlibrary{calc,arrows.meta,positioning,decorations.pathreplacing}

\usepackage{microtype}

\usetikzlibrary{arrows.meta}

\makeatletter \def\switch@array{}\makeatother

\theoremstyle{plain}
  \newtheorem{theorem}{Theorem}[section]

\theoremstyle{definition}

\theoremstyle{remark}

  \newtheorem{example}[theorem]{Example}

\numberwithin{equation}{section}

  \let\de=\delta 
\let\ve=\varepsilon  \let\ga=\gamma 
  \let\om=\omega 
\let\si=\sigma

\newcommand{\opunit}{\text{1}\kern-0.22em\text{l}}

\DeclareMathAlphabet{\mathpzc}{OT1}{pzc}{m}{it}

\newcommand{\fig}{Fig.}

\newcommand{\id}{\textrm{d}}

\usepackage{floatrow}
\usepackage{caption}
\DeclareCaptionJustification{justified}{\leftskip=0pt \rightskip=0pt \parfillskip=0pt plus 1fil}

\numberwithin{equation}{section}

\usepackage{xcolor,soul}
\newcommand{\tp}{\textcolor{purple}}
\newcommand{\tb}{\textcolor{blue}}
\newcommand{\tg}{\textcolor{teal}}

\newcommand{\tr}{\textcolor{red}}

\begin{document}

\title{Bringing calorimetry (back) to life} 

\author{
Faezeh Khodabandehlou$^{1}$, Christian Maes$^{1}$, and \'Edgar Rold\'an$^{2}$\\[1ex]
\small $^{1}$Department of Physics and Astronomy, KU Leuven, Belgium\\
\small $^{2}$ICTP -- The Abdus Salam International Centre for Theoretical Physics, Trieste, Italy
}

\begin{abstract}
Micro-calorimetry offers significant potential as a quantitative method for studying the structure and function of biological systems, for instance, by probing the excess heat released by  cellular  or sub-cellular structures, isothermal or not, when external parameters change. 
We present the conceptual framework of nonequilibrium calorimetry, and as illustrations, we compute the heat capacity of biophysical models with few degrees of freedom related to ciliar motion (rowing model) and molecular motor motion (flashing ratchets). Our quantitative predictions    reveal intriguing dependencies of the  (nonequilibrium) heat capacity as a function of  relevant biophysical parameters, which can even take negative values as a result of biological activity.
\end{abstract}

\maketitle

\section{Introduction}
Biological heat experiments pioneered calorimetry. Most influential were the famous Lavoisier-Laplace experiments (1780) aimed at understanding the respiratory system of living systems. They demonstrated that animal respiration is a form of  combustion (chemical oxidation), and (biological) heat (production)  became since then a measurable physical quantity.  Yet, for general physics, the constructive power of calorimetry showed most vividly around the birth of quantum mechanics.  In 1911 Einstein proposed to solve ``the most important problem in kinetic theory'' ({\it i.e.}, the one of specific heats) by introducing the Planck hypothesis in condensed matter physics as well. The experiments of Nernst and others led to the formulation of the Third Law of thermodynamics and the powerful tool of thermal response was developed for both theoretical and experimental low-temperature physics~\cite{EnW}.  The first applications of calorimetry to soft-matter systems date from the mid-20th century, in particular by the interest in polymers, liquid crystals, and biological membranes.  For example, in the 1940-1950s, differential scanning calorimetry was applied to study glass transitions and melting behaviors in polyethylene and polystyrene, which was an important discovery for plastics and rubber technologies.  In the 1960s came phase transitions in liquid crystals, and calorimetry could pick up small heat changes indicating reordering of molecules.  For the study of cell membranes, important work included \cite{blume1988applications,BLUME1985473,chapman1968biological}, where gel-to-liquid crystalline transitions were observed in lipids.  In all these efforts, calorimetry could detect transitions even when there were few other indicators.
Turning into the 21st century, successes continued in soft-matter physics and physical chemistry~\cite{jeong2001modern,saitta2022calorimetric}, while, more than two centuries after Lavoisier-Laplace, systematic experimental studies have re-appeared measuring heat produced by biological systems.  Including more recent work, we refer, for example, to~\cite{fessas2017isothermal,Arunachalam2023, Rodenfels2019, Hong2020} which reported rates of heat dissipation by single cells to be within orders of magnitude ranging $(0.1 n \text{W}-\mu\text{W})$ with the lowest value being near the finest (sub-nanowatt) sensitivity limits of microcalorimeters~\cite{Hong2020}.

However,  heat exchange for driven or active systems, like studying calorimetry for living systems (and for steady nonequilibrium systems more generally), requires a new framework,  not just a new technique.  The main input is the notion of {\em excess} heat, defined in Refs.~\cite{oono,HS}, and exploited recently in active and complex systems in Refs.~\cite{activePritha,wang2015landscape,datta2022second,manzano2024thermodynamics}. A system in a steady nonequilibrium condition dissipates heat into a thermal bath kept at some fixed temperature.  That is called the {\em house--keeping} heat; there is, on average, a constant heat power (Joule heating) as for every steady dissipative system~\cite{oono,HS}.  The condition of the system is kept steady indeed while irreversible work is done on the system, be it in terms of food, internal fueling, or external fields.  Yet, as some external or internal parameter is slowly changed, be it the temperature of the heat bath making the cellular environment, or the load of a molecular motor, there appears an extra or excess heat entirely due to that change.  That is our main interest: how the excess heat depends on physical parameters.
For example, as shown in Refs.~\cite{jir,epl,Pe_2012,jchemphys,calo}, it is possible to extract the heat capacity of a nonequilibrium system from the out-of-phase component of the (extra) heat dissipated as a result of a periodic  modulation of the temperature of the thermal bath within which the nonequilibrium system is embedded.  This approach focuses on understanding the response of nonequilibrium systems (biological or not) to perturbations (of temperature or other quantities), and is recently gaining considerable momentum~\cite{calo,datta2022second,fernandes2023topologically}  as a research avenue providing novel insights beyond the already existing extensive research on steady-state properties of active systems~\cite{wang2015landscape,barato2015thermodynamic,fodor2016far,horowitz2020thermodynamic,yang2021physical,roldan2021quantifying,loos2023measurement,roldan2024thermodynamic}, going beyond diagnozing the violation of standard fluctuation-dissipation relations~\cite{martin2001comparison,mizuno2007nonequilibrium,gnesotto2018broken}, 

 For more background and derivations of nonequilibrium calorimetry, we refer to Refs.~\cite{jir,epl,Pe_2012,jchemphys,calo}, where a plethora of illustrative examples of nonequilibrium stochastic models are discussed in detail. Here, instead, we focus on  applications of nonequilibrium calorimetry in realistic models borrowed from theoretical biophysics.  
While this can be seen as a continuation of the project and the explorations in~\cite{activePritha}, in the present paper we take a physics-of-life perspective as we provide quantitative predictions for the order of magnitude of the nonequilibrium heat capacity and how it depends on biophysical parameters in life processes at the microscale related to ciliar motion and molecular motor stepping. 
Unlike in previous  analyses where the heat capacity (a thermodynamic response function to temperature) was mostly studied as a function of the temperature itself, here we rely on a more common biophysical scenario of constant temperature and study the (nonequilibrium) heat capacity {\em at room temperature} as a function of e.g. the rate of ATP hydrolysis, the value external forces, etc.\\

\underline{Plan of the paper:}  We introduce the basic concepts on nonequilibrium calorimetry  in Sec.~\ref{sec:2new}. The main ingredient is heat flux (dissipated power), and we show how AC-calorimetry delivers the (nonequilibrium) heat capacity when the system is time-inhomogeneous.  We also present a second, more theoretical method, using the quasipotential (see Appendix~\ref{sec:2} for details).  We apply our methodology to theoretical   models of nonequilibrium biological fluctuations in Sec.~\ref{sec:cilium} and Sec.~\ref{sec:motor}.  We present both continuous (Langevin) and discrete (Markov jump) models for cilium motion (Sec.~\ref{sec:cilium}) and for molecular motor motion (Sec.~\ref{sec:motor}), and each time, we obtain the heat capacities for biophysically-relevant parameters. Concluding remarks are discussed in Sec.~\ref{sec:conc}, and additional points and results are provided in the Appendices.

\section{Nonequilibrium calorimetry for physicists of life}
\label{sec:2new}
 
We provide the essentials of nonequilibrium calorimetry for application in biophysics.  We focus on how the (nonequilibrium) heat capacity can be retrieved from the response of a biological system to temperature changes. Here we outline the key concepts and techniques, while we relegate detailed physical definitions and mathematical derivations to Appendix~\ref{sec:2}.

\subsection{Definition of nonequilibrium heat capacity}

The main idea behind nonequilibrium calorimetry follows from extending ideas from classical `textbook'  macroscopic equilibrium thermodynamics  to the physics of open steady out-of-equilibrium systems in contact with a heat bath like found in living systems. In particular, we focus on  systems that,  when the temperature of the bath is $T$, dissipate on average a positive steady heat flow $\dot{\mathcal{Q}}_T$ to the bath due to, e.g., active processes such as chemical fuel consumption. This steady flow is often referred to as {\em housekeeping}.

Throughout this work, we use the convention by which $\cal{Q}$ stands for  the heat {\em dissipated} to the bath: $\cal{Q}>0$ positive means it is dissipated from the system to the bath, while for $\cal{Q}<0$ heat is absorbed by the system from the bath. The second law of thermodynamics implies that all nonequilibrium systems dissipate a steady heat flow, i.e. $\dot{\mathcal{Q}}_T\geq 0$, with equality when the system is in equilibrium. The response of such nonequilibrium systems to time-dependent perturbations is highly nontrivial. For instance, for a  time-dependent stimulus (e.g. a sudden quench of the bath temperature) heat may either be dissipated or absorbed by the system, as we show below with examples. 

\begin{figure}[H]
\centering

\begin{subfigure}[t]{0.48\textwidth}
\centering
\begin{tikzpicture}
\begin{axis}[
    width=\linewidth,
    height=4cm,
    axis lines=left,
    xlabel={Time (s)},
    xlabel style={xshift=2cm, anchor=north},
    ylabel={Temperature ($^\circ$C)},
    ylabel style={
        at={(axis description cs:0.2,1)}, 
        anchor=south,
        rotate=-90                            
    },
    xmin=0, xmax=10,
    ymin=23.5, ymax=26.5,
    xtick=\empty,
    ytick={24,25,26}
]
\addplot[thick,blue] coordinates {
    (0,25) (4,25) (4,26) (10,26)
};
\addplot[dashed, thick, black]
coordinates {(4,23.5) (4,26.5)};
\node[left] at (axis cs:2,25.3) {$T$};
\node[right] at (axis cs:6,26.3) {$T+\mathrm{d}T$};
\draw[<->,thick] (axis cs:4.3,25) -- (axis cs:4.3,26) 
    node[midway,right] {$\mathrm{d}T$};
\end{axis}
\end{tikzpicture}

\vspace{0.3cm}

\begin{tikzpicture}
\begin{axis}[
    width=\linewidth,
    height=4cm,
    axis lines=left,
    xlabel={Time (s)},
    xlabel style={xshift=2cm, anchor=north},
    ylabel={$\dot Q$},
        ylabel style={
        at={(axis description cs:-0.08,0.9)}, 
        anchor=south,
        rotate=-90                            
    },
    xmin=0, xmax=10,
    ymin=0, ymax=1.5,
    xtick=\empty,
    ytick=\empty
]
\addplot[dashed, thick, black]
coordinates {(4,0) (4,1.5)};
\addplot[thick, purple]
coordinates {(6.66,0.9) (10,0.9)};
\addplot[thick, purple]
coordinates {(0,0.5) (4,0.5)};
\addplot[
    thick,
    purple,
    samples=400,
    domain=4:6.66,
    name path=curve
]
{
ifthenelse(
    x < 6.15,
    0.5 + 0.9*sin(deg(2*pi*0.2*(x-4))),
    0.9 + 0.15*cos(deg(2*pi*1*(x-5.9)))
)
};

\addplot[
    dashed,
    thick,
    purple,
    domain=4:6.66,
    name path=baseline
]
{0.9};

\addplot[
    fill=purple!40,
    fill opacity=0.4
]
fill between[
    of=curve and baseline
];

\node[left] at (axis cs:2.5,0.7) {$\dot Q_T$};
\node[right] at (axis cs:6.5,1.2) {$\dot Q_{T+\mathrm{d}T}$};
\node at (axis cs:5.4,0.6) {$\delta Q_{\text{exc}}$};
\addplot[dashed, thick, purple]
coordinates {(4,0.9) (6.66,0.9)};
\end{axis}
\end{tikzpicture}
\caption{}
\end{subfigure}
\hfill
\begin{subfigure}[t]{0.48\textwidth}
\centering
\begin{tikzpicture}
\begin{axis}[
    width=\linewidth,
    height=4cm,
    axis lines=left,
    xlabel={Time (s)},
    xlabel style={xshift=2cm, anchor=north},
    ylabel={Temperature ($^\circ$C)},
       ylabel style={
        at={(axis description cs:0.2,1)},
        anchor=south,
        rotate=-90                       
    },
    xmin=0, xmax=10,
    ymin=23.5, ymax=26.5,
    xtick=\empty,
    ytick={24,25,26}
]
\addplot[thick,blue,samples=200,domain=0:8.8] 
{ifthenelse(x<4,25, 25 + sin(deg(2*pi*0.6*(x-4))))};
\node[right] at (axis cs:6.4,25.7) {$\frac{1}{\omega_b}$};
\node[left] at (axis cs:2,25.3) {$T$};
\draw[<->,thick] (axis cs:6.1,26.2) -- (axis cs:7.8,26.2) 
    node[midway,right]{};
\node[right] at (axis cs:8.1,25) {$2\,\delta T$};
\draw[<->,thick] (axis cs:9.6,24) -- (axis cs:9.6,26) 
    node[midway,right]{};
\end{axis}
\end{tikzpicture}

\vspace{0.3cm}

\begin{tikzpicture}
\begin{axis}[
    width=\linewidth,
    height=4cm,
    axis lines=left,
     clip=false,
    xlabel={Time (s)},
    xlabel style={xshift=2cm, anchor=north},
    ylabel={$\dot Q$},
    ylabel style={
        at={(axis description cs:-0.08,0.9)},
        anchor=south,
        rotate=-90
    },
    xmin=0, xmax=10,
    ymin=-0.6, ymax=1.2,
    xtick=\empty,
    ytick=\empty
]

\addplot[thick,purple,domain=0:4] {0.4};

\addplot[
    thick,
    purple,
    samples=200,
    domain=4:9.3,
    xshift=3mm
]
{0.4 + 0.5*sin(deg(2*pi*0.6*(x-4)))};

\addplot[dashed,gray] coordinates {(4.3,0.4) (10,0.4)};
\draw[line width=0.8pt]
  ($(axis cs:4,0.4)+(-1mm,1.2mm)$) -- ++(2mm,-2mm)
  ($(axis cs:4,0.4)+( 1mm,1.2mm)$) -- ++(2mm,-2mm);

\draw[line width=0.8pt]
  ($(axis cs:4,-0.6)+(-1mm,1.2mm)$) -- ++(2mm,-2mm)
  ($(axis cs:4,-0.6)+( 1mm,1.2mm)$) -- ++(2mm,-2mm);

\node[left] at (axis cs:2.5,0.6) {$\dot Q_T$};

\end{axis}
\end{tikzpicture}

\caption{}
\end{subfigure}
\caption{{\bf Nonequilibrium calorimetry at a glance.} (a) Step excitation: temperature (top) and heat flow (bottom) as a function of time.  Illustration of the released power (= heat flux) when the bath temperature is changed $T\to T+\text{d}T$, with $T=25^o$C and d$T=1^o$C in this example.  The excess heat $\delta \cal Q^{\text{exc}}$ due to that change in temperature equals the shaded area, and the nonequilibrium heat capacity equals $C_T= -\delta \cal Q^{\text{exc}}/\text{d}T$. (b) Oscillatory (AC) excitation: temperature (top) and heat flow (bottom); see text for further details. The excess heat and nonequilibrium heat capacity are estimated respectively; see also Fig.~\ref{comp} for a specific illustration in a stochastic model of ciliar motion.}
\label{fig:temp_heat_2x2}

\end{figure}
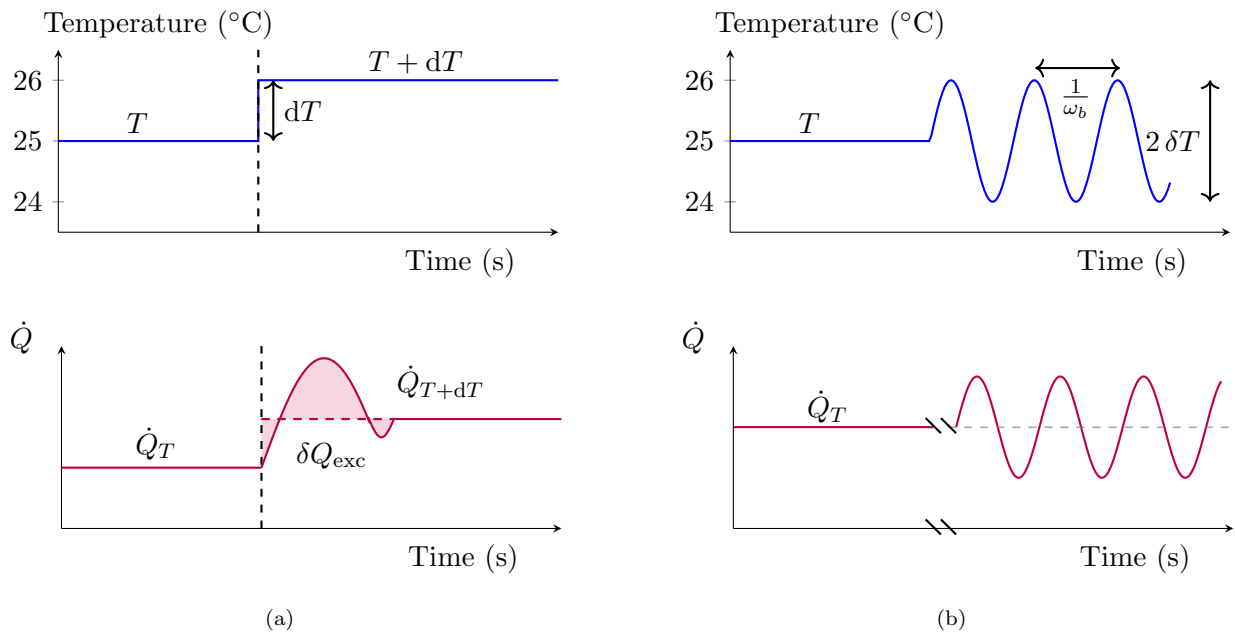

To be more specific, we suppose that the bath's temperature changes from $T$  to $T+\id T$, after which it remains equal to $T+\id T$.  That change in itself creates excess heat (or in defect) compared to the stationary power dissipated at temperature $T+dT$.  
\\ Formally, the so-called {\em excess heat} following the step excitation $T\to T+\text{d}T$ is  defined as
\begin{equation}
\delta Q_{\text{exc}}= \int_0^\infty \left[ \dot{Q}(t) - \dot{Q}_{T+\text{d}T}\right] \text{d}t ,\label{eq:1}
\end{equation}
i.e., it is the heat  $\dot{Q}_{T+\text{d}T}$ to the thermal bath, renormalized to the new steady-state value.
In other words,  $\delta Q_{\text{exc}}$ can be represented as the net area swept (after the step excitation $T\to T+\text{d}T$)  by the instantaneous heat flux over time in excess to  the final steady value   $\dot{Q}_{T+\text{d}T}$; see Fig.~\ref{fig:temp_heat_2x2}(a).

 The nonequilibrium heat capacity at temperature $T$ is defined in terms of $-\delta{\cal Q}^\text{exc}$, the excess heat uptake by the system from the bath:
\begin{equation}\label{cdef}
C_T = -\frac{\delta {\cal Q}^\text{exc}}{\id T} .
\end{equation}
Notably, it was shown that $\delta {\cal Q}^\text{exc}$ can take negative (excess heat dissipated to the bath) but also positive (excess heat absorbed by the system) values, the latter being genuine to systems out of equilibrium; it is also reported here for molecular motor motion (see Sec.~\ref{sec:motor}). 
 In equilibrium, the housekeeping heat $\dot{\cal{Q}}_T=0$ at all temperatures $T$,  and the excess heat $\delta Q_{\text{exc}}=\delta Q$ equals the total heat, leading to the standard equilibrium definition of the heat capacity, $C_T = -\delta {\cal Q}/\id T $, obeying $C_T \geq 0$ by virtue of thermodynamic stability.
 
 In this work we do the first steps towards analyzing how nonequilibrium thermal response reflects biological functioning, studying how  $C_T$ depends on relevant biophysical parameters.  The temperature-dependence of $C_T$ is less relevant here, and we typically consider values at or around room temperature. Similarly, one could vary other parameters $\lambda$ than the temperature in~\eqref{cdef}, and be interested in other (isothermal) response coefficients with respect to small changes in $\lambda$; those will not be considered in the present paper.

\subsection{Estimating the nonequilibrium heat capacity}
Definition \eqref{cdef} needs some rewiting to become more practical and ready to be put at work with experimental and/or numerical data.  Experimentally, estimating the nonequilibrium heat capacity would require accurate measurements of the excess heat, an enterprise that we expect could be tackled with modern micro-calorimeters through e.g. differential scanning calorimetry~\cite{fessas2017isothermal,Arunachalam2023, Rodenfels2019, Hong2020}.  From the theoretical perspective,   two  computational schemes for obtaining the nonequilibrium heat capacity have been introduced; see also \cite{activePritha, epl, calo,jchemphys,Pe_2012}. One relies on evaluating (preferably analytically) the temperature dependency of the so-called quasipotential  $V$ (see Appendix~\ref{M1}), which allows one to access directly the heat capacity from the step excitation given by Eq.~\eqref{cdef}. This approach is often more suitable for Markov-jump models used to model e.g. molecular stepping processes, but technically challenging for continuous models (e.g. Langevin dynamics). For the latter, a more suitable method is found in so-called AC-calorimetry (see Appendix~\ref{M1} and below), which gives access to an {\em estimate} of $C_T$ in Eq.~\eqref{cdef} by analyzing the response of the heat flux to a oscillatory temperature excitation.  

The method of AC--calorimetry  for nonequilibrium systems  is illustrated in Fig.~\ref{fig:temp_heat_2x2}(b). Following Ref.~\cite{calo}, the main idea is again to analyze the thermal response of the system that is initially at temperature $T$ dissipating a net heat flow $\dot{\cal{Q}}_T$. 
Next, the bath temperature $T_b$ is modulated periodically in time with (small) frequency $\omega_b>0$, {\it e.g.}, following a sinusoidal modulation with (small) relative amplitude $\epsilon_b $ around $T$, in other words
\begin{equation}\label{eq:TperMaes}
T_b(t)=T[1-\epsilon_b\sin(\omega_b t)].  
\end{equation}
During the `AC' sinusoidal modulation, we measure the time-dependent average heat flux, $\dot{\cal Q}(t)$. 
For time-periodic temperatures following Eq.~\eqref{eq:TperMaes}, in  linear order and neglecting $O(\omega_b^2,\epsilon_b^2)$, $\dot{\cal Q}(t)$ can be approximated as
\begin{equation}\label{mama}
   \dot{\cal Q}(t) \simeq  \dot{\cal Q}_T +\epsilon_b\, T\,[B_T\, \sin(\omega_b t) + C_T\,\omega_b\, \cos(\omega_b t)],
\end{equation}
Here, the first Fourier coefficient retrieves the mean stationary (housekeeping) heat flux rate $\dot{\cal{Q}}_T$ before the AC driving starts,  $B_T\delta T =(\dot{\cal Q}_{T+ \delta T} - \dot{\cal Q}_T )/\epsilon_b T$, while $C_T$ is the nonequilibrium heat capacity as in \eqref{cdef}; see Fig.~\ref{comp} 
and Appendix~\ref{sec:2} for further details.



To demonstrate how to put these ideas to work, in the following sections we present illustrative example applications to theoretical biophysics  models that capture essential features of some biological functioning. 

\section{Heat capacity of ciliar beating}
\label{sec:cilium}

The ``rower model'' is a minimal theoretical framework developed to study the coordinated beating of cilia through hydrodynamic interactions. In this model, each cilium is represented as a two-state oscillator that switches direction upon reaching certain thresholds, capturing the essential features of ciliary motion.

It was introduced in \cite{cosentino2001rowers}
where the ``rowers'' are simplified oscillators that interact solely through hydrodynamic forces. 
In what follows, we ignore the hydrodynamic interactions (possibly leading to coordinated motion) and we consider a model for a single-cilium rower.\\
We compute its heat capacity $C$ as a function of structural parameters. 
Let us start with an equilibrium estimate:  a cilium is like a flexible elastic rod of length $\sim 5-10 \mu$m, and diameter $\sim 0.2 \mu$m.
    If we model it (passively) with Fourier modes along its length, some 10–100 bending modes are relevant.
Since each mode is a degree of freedom, we end up with a rough (equilibrium) estimate of $C^\text{eq} \simeq 10-100 \,k_B = (10^{-23}-10^{-21})$ J/K$= (0.01-1)$zJ/K.

\subsection{Nonequilibrium calorimetry of  ciliar beating I: Langevin  (diffusion)  model}
\begin{figure}[H]
	\centering
	\begin{subfigure}{0.27\textwidth}
		\centering
		\includegraphics[width=1\linewidth]{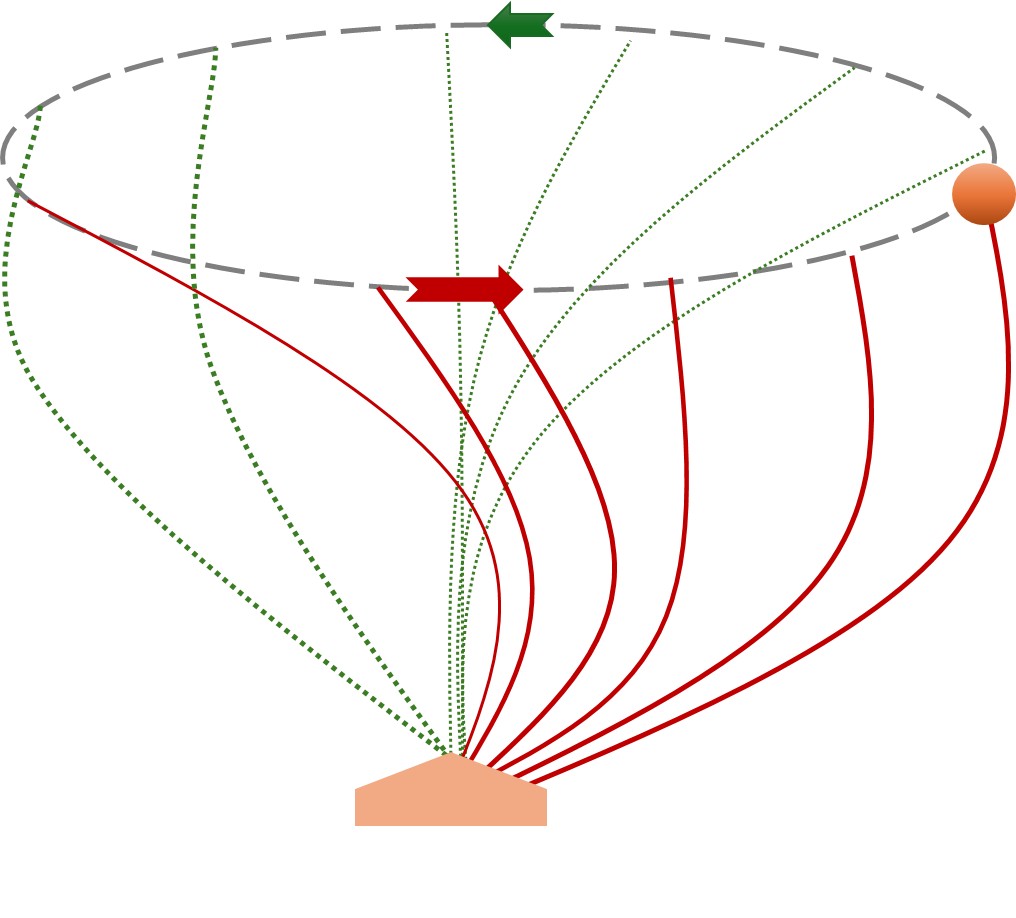}
		\caption{}
	\end{subfigure}
	\hspace{0.5em} 
	\begin{subfigure}{0.3\textwidth}
		\centering
		\includegraphics[width=1.05\linewidth]{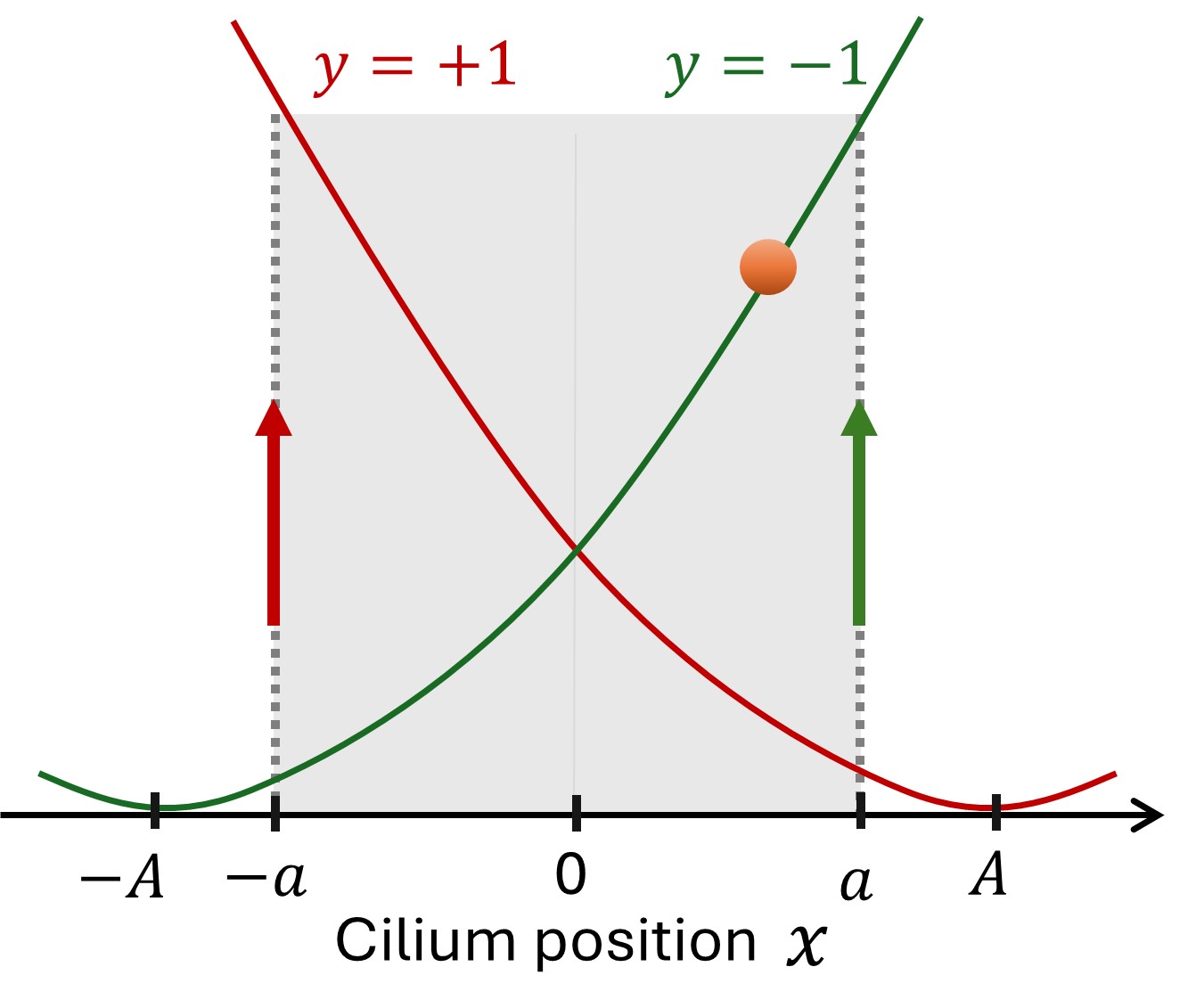}
		\caption{}
	\end{subfigure}
	\hspace{1em} 
	\begin{subfigure}{0.36\textwidth}
		\centering
		\includegraphics[width=\linewidth]{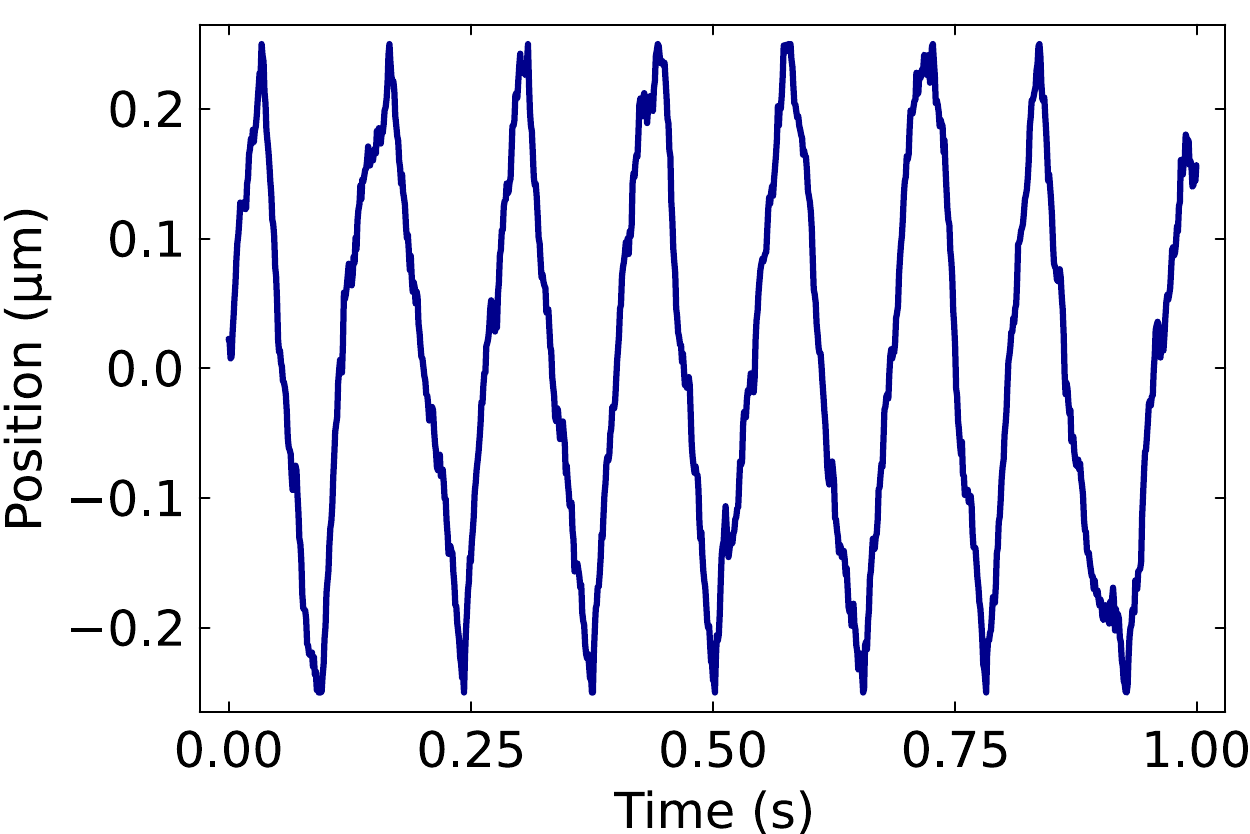}
		\caption{}
	\end{subfigure}
	\caption{{\bf Diffusive rower model for a single cilium.} (a) Sketch of the motion of a single cilium whose tip motion is probed by a bead (orange circle). (b) Sketch of the stochastic oscillation mechanism of the rower model described in \eqref{rowm}: the tip of the cilium (orange circle) diffuses in a harmonic potential centered in $-A$ (green line) until it reaches  $-a>-A$, after which the potential's minimum is immediatly switched to $A$ (red line) until the next passage through $a$ etc. (c) Numerical simulation for the cilium's  trajectory over one second for bath temperature $ \, T=300 \,\mathrm{K}$, friction coefficient  $\gamma = 0.2\,\text{pN s}/\mu\text{m}$, stiffness $\kappa = 1.5\,\text{pN}/\mu\text{m}$, and length parameters $A = 1\,\mu\text{m}$ and $a = 0.25\,\mu\text{m}$; see also Ref.~\cite{gupta2025}.}
	\label{rowingModeltra}
\end{figure}

To describe positional cilia movement in a viscous fluid, we use the model described in \cite{kotar2010hydrodynamic,bruot2016realizing,gupta2025}.
To this aim, we consider the one-dimensional projection of the motion of the tip of a single cilium  whose basal body is attached to a substrate; see Fig.~\ref{rowingModeltra}(a).  The tip of the cilium  beats rhythmically in a fluid, displaying oscillating  fluctuations near  $-a$ and $a$, with $a\geq 0$. The movement of the tip position along the horizontal axis  is analogous to that of a spring which switches between two harmonic potentials depending on its position,
\begin{equation}
U(x,y)=\frac{1}{2} \, \kappa\, (x-y A)^2,\qquad y=\pm 1,
\end{equation}
where $\kappa>0$ denotes the stiffness, and $2A $ is the separation distance between the minima of the two potentials. The state of the tip is characterized by two degrees of freedom, the tip position $x$, and $y= \pm1$ indicating which potential is currently active. In equilibrium, where $A=0$, the particle displays equilibrium fluctuations around $x=0$. However, for the  nonequilibrium rowing model, $A > a$. 
The tip diffuses  in  the potential $\kappa(x-A)^2/2$ (or, $\kappa(x+A)^2/2$) and begins to move toward $+a$ (resp. $-a$) when  $y = +1$ ($y = -1$) until it crosses $-a$ ($+a$).  When the potential switches $y \rightarrow -y$, the net drift of the tip's motion is reversed, yet the motion is restricted to the interval $[-a, a]$ (with $A > a$, so the minimum is not reached).
Each potential switch  injects energy into the system, causing the bead to reverse the net direction and resulting in sustained oscillations that resemble the rhythmic beating of a cilium; see \fig~\ref{rowingModeltra}(a) and  (b). Notably, for any $a<A$,  the  switches $y\to -y$ occur at a finite (nonzero) rate that equals the inverse of the mean first-passage time to $\pm a$ starting from $\mp a$ in potential $U(x,\pm 1)$. Such unidirectional transitions  break the condition of detailed balance, leading to a nonequilibrium stationary state in the long-time limit.  In  Section \ref{dis}, we present a less conventional, yet analogous and illuminating discrete rowing model described as a Markov-jump process with the advantage of satisfying local detailed balance \cite{ldb}, and thus allowing a clean equilibrium limit.

Because the system is small and immersed in a (viscous) fluid, the tip is also influenced by Brownian motion in the form of thermal (Gaussian white) noise. The result is a diffusion process for the tip (or bead attached to it) interrupted by jumps.  Altogether, the bead's motion is governed by the overdamped Langevin equation,
\begin{eqnarray}
\gamma\dot{x}_t& =& 
 - \kappa (x_t-y_t A) + \sqrt{2 k_B T \gamma }\,\,\xi_t, \label{rowm}
\end{eqnarray}
where we need to remember the jumps (switches) associated with the process $y_t$.   The drag-force coefficient is $\gamma= 6\pi\eta r$  (with $\eta$ the fluid viscosity, and  $r$ the bead radius), and $\xi_t$ denotes a standard Gaussian white noise.  
An example  trajectory of the tip is given in \fig~\ref{rowingModeltra} (c), showing the characteristic beating  motion.

\begin{figure}[h]
 \centering
      \begin{subfigure}{0.49\textwidth}
         \centering
         \def\svgwidth{\linewidth}        
        \includegraphics[scale = 0.77]{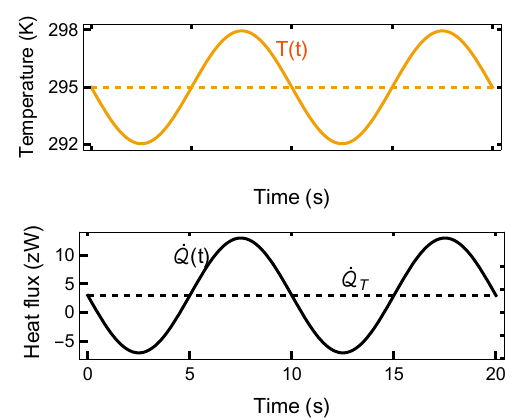}
        \caption{}
     \end{subfigure}
     \hfill
     \centering
      \begin{subfigure}{0.49\textwidth}
         \centering
         \def\svgwidth{0.8\linewidth}        
        \includegraphics[scale = 0.85]{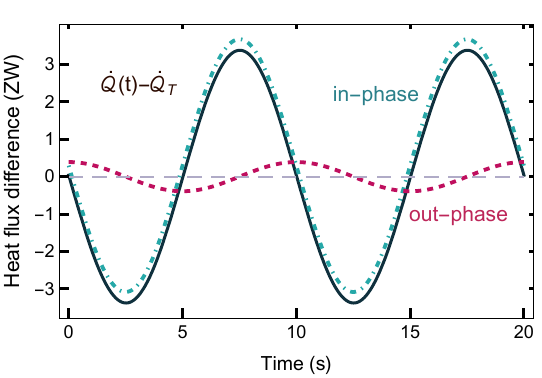}
        \caption{}
     \end{subfigure}
\caption{{\bf Nonequilibrium (AC-) calorimetry at work: rower model for ciliar motion.} \small{(a) Top: heat-bath temperature oscillations $T_b(t)=T[1-\epsilon_b\sin(\omega_b t)]$   used for AC-calorimetry. Bottom: AC and DC expected heat fluxes.  (b) The differences in the expected heat fluxes $\dot{\mathcal{Q}}(t) - \dot{\mathcal{Q}}_T$ in Eq.~\eqref{mama} specifically for the stochastic nonequilibrium dynamics associated with the rower model of a single cilium motion; see Eq.~\eqref{rowm}. The data are shown for reference temperature $T = 295\,\text{K}$,  temperature relative amplitude $\epsilon_b = 0.01$, temperature oscillation frequency $\omega_b= (\pi/5)$Hz. For  the sake of visualization, the nonequilibrium heat capacity $C_T$ is taken in this plot to be ten times larger than its actual value. The  in-phase Fourier component is $B_T\epsilon T \sin(\omega t)$ while the  out-of-phase  component is $C_T \epsilon_b T \omega_b \cos(\omega_b t)$. Here,  $C_T = 0.15\,k_B$, friction coefficient  $\gamma = 0.2\,\text{pN s}/\mu\text{m}$, stiffness $\kappa = 1.5\,\text{pN}/\mu\text{m}$, and length parameters $A = 1\,\mu\text{m}$ and $a = 0.25\,\mu\text{m}$. 
}}\label{comp}
\end{figure}

Possible measurement of the dissipated power proceeds via active microrheology with external forcing: one attaches a microscopic  bead to the tip of the cilium, by which displacements of the cilium in the nanometer range may be measured experimentally. From the bead fluctuations,  the  heat capacity may be estimated via  AC-calorimetry. In particular, for the time-periodic temperature modulation \eqref{eq:TperMaes} the expected heat flux at time $t$ reads
\begin{eqnarray}\label{powerrowing}
   \dot{\cal {Q}}(t) &=& \mu \kappa\,\left[ \kappa \left\langle (x_t - y_t A)^2\right\rangle_t-  k_B T_b(t)  \right] , 
\end{eqnarray}
 For this example, $\dot{\cal {Q}}(t)$ is given by  the inverse of the relaxation time $\tau_r=(\mu\kappa)^{-1}$  of the position in the harmonic potential $\kappa (x-y A)^2/2$ times a  term that vanishes  when an effective equipartition theorem holds at all times $t$ during the driving. 

 \begin{figure}[h]
 \centering
      \begin{subfigure}{0.49\textwidth}
         \centering
         \def\svgwidth{0.8\linewidth}        
        \includegraphics[scale = 0.38]{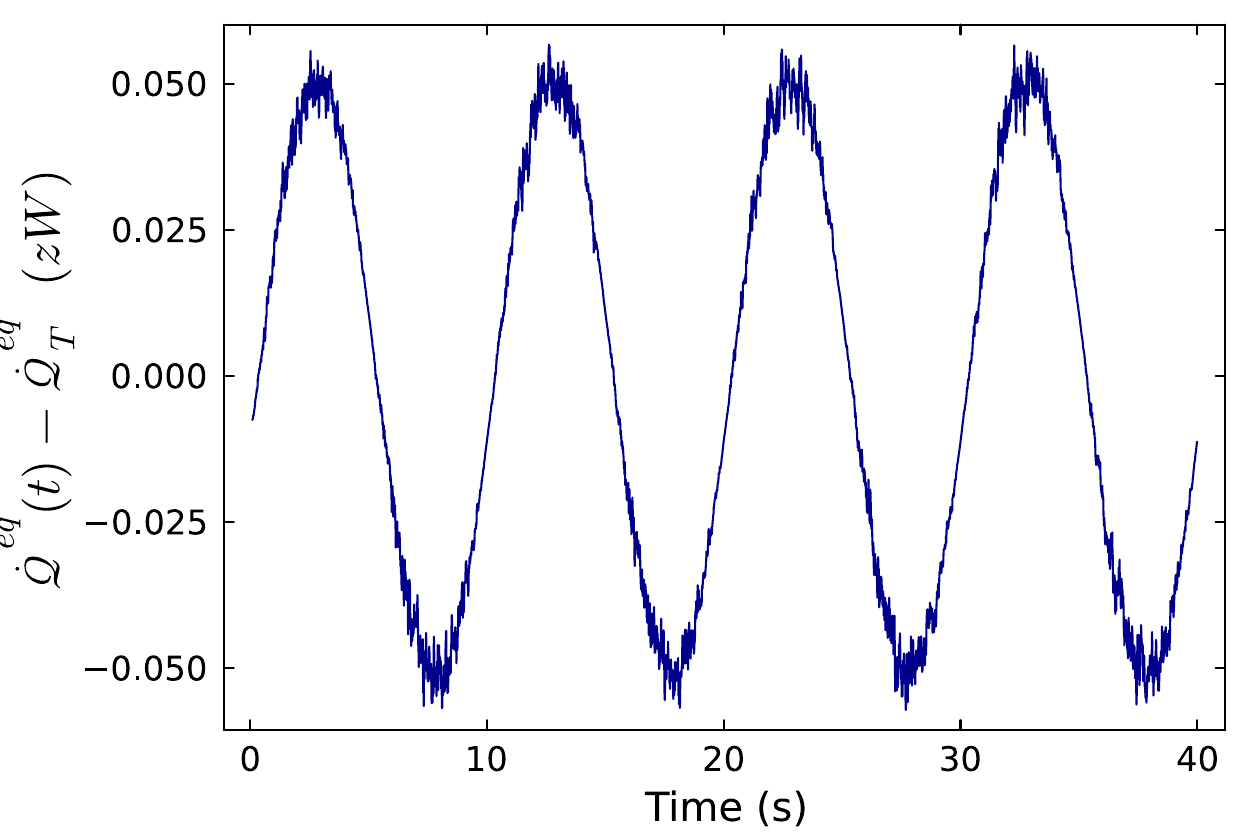}
        \caption{$\,$} \label{hcgamma0a}
     \end{subfigure}
     \hfill
     \centering
      \begin{subfigure}{0.49\textwidth}
         \centering
         \def\svgwidth{0.8\linewidth}        
        \includegraphics[scale = 0.38]{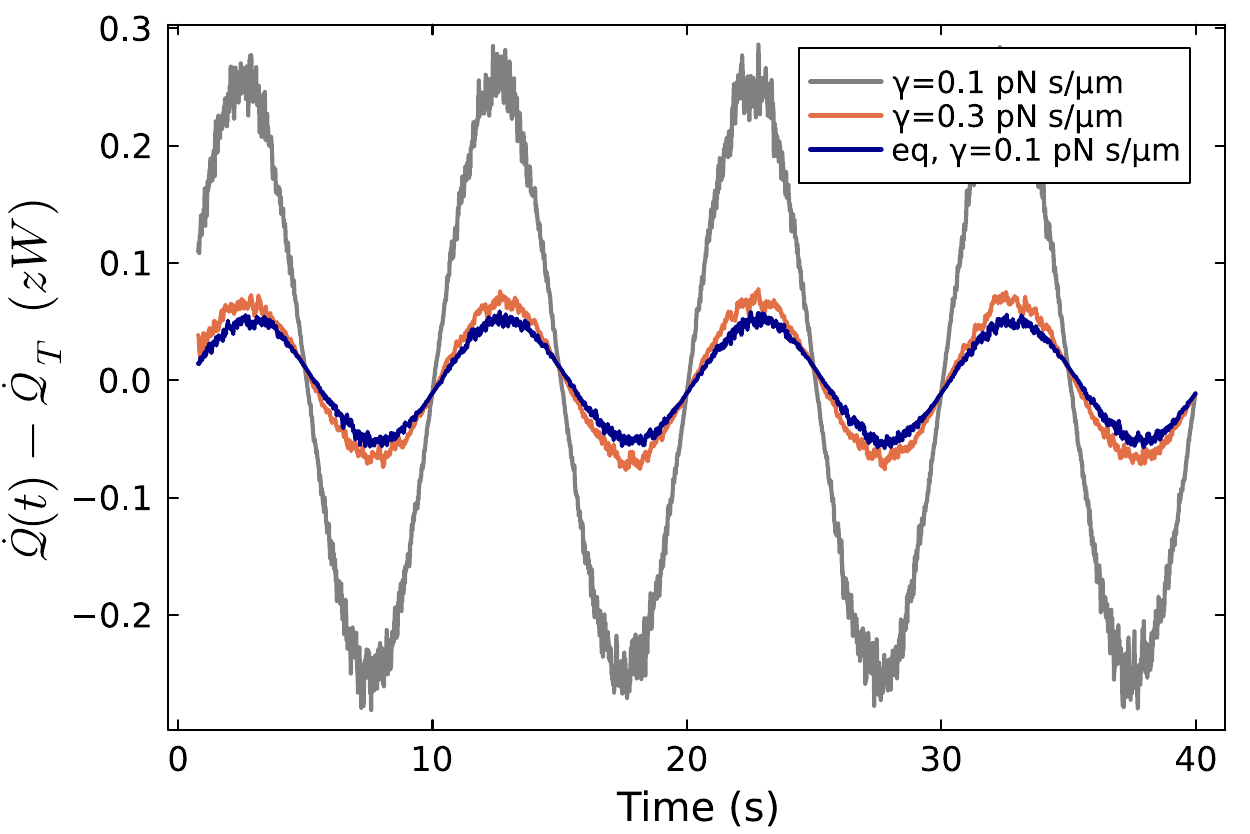}
        \caption{$\,$}\label{hcgamma0b}
     \end{subfigure}
\caption{{\bf{Expected excess heat flux as a function of time for the diffusive rower obtained from AC calorimetry (a) in equilibrium, and (b)  out of equilibrium.}} \small{Estimates from numerical simulations of the time-dependent (expected) excess heat flux, obtained as the difference between AC (total $   \dot{\cal {Q}}(t) $)  and DC (housekeeping $\dot{\cal Q}_T$) expected heat fluxes as a function of time for the rowing model for parameter values $\kappa = 1.5 \mathrm{pN/\mu m}$,  $a = 0.25 \mu\mathrm{m}$, and temperature modulation $T_b(t)= (1-\epsilon_b\,\sin(\omega_b t))T$, with $T=300$K, $\epsilon_b=0.01$, $\omega_b= \pi/5 $Hz.  For the equilibrium case (a) $A=0$ and  $\gamma= 0.1$\,pN\,s/$\mu$m, while for the nonequilibrium cases (b)  $A = 0.5  \mu\mathrm{m}$ and $\gamma$ as specified in the legend, where we also added the equilibrium case at $\gamma= 0.1$\,pN\,s/$\mu$m for comparison. All results are averages over $10^5$ numerical simulations.  }}  
\end{figure}
  
 Fig.~\ref{hcgamma0a} shows the expected heat flux associated with  equilibrium  fluctuations in the rower model (setting $A=0$) as a function of time for a temperature modulation given by  $T_b(t)= (1-\epsilon_b\,\sin(\omega_b t))T$ [Eq.~\eqref{eq:TperMaes}] with mean value $T=300$K, amplitude $\epsilon_b=0.01$ around the room temperature, and frequency $\omega_b=(\pi/5)$Hz. This modulation corresponds to experimentally-reasonable temperature oscillations of amplitude $\epsilon_b T=3^{o}$C, and yield heat flux oscillations of the order of yoctowatts ($10^{-24}$W) which is 15 orders of magnitude smaller than some of the most accurate experimental measurements of heat fluxes so far in experimental  calorimetric devices. From these oscillations, we estimate $C_T^{\text{eq}} = k_B/2$ in equilibrium, as expected for a diffusion in a harmonic potential.

 To calculate the nonequilibrium heat capacity via AC-calorimetry, we first repeat the same analysis carried for equilibrium dynamics ($A=0$) but for $A>0$ using the same temperature driving and exploring different values of the friction coefficient $\gamma$.   Values of the expected excess heat fluxes as a function of time are reported in Fig.~\ref{hcgamma0b}.
In the spirit of Eq.~\eqref{eq:1}, the expected excess heat fluxes are obtained by subtracting the (housekeeping) component $\dot{\cal{Q}}_T$ at temperature $T$ from the total expected heat flux under AC-modulation $\dot{\cal{Q}}(t)$.  Figure~\ref{hcgamma0b} shows that the expected excess heat under AC-modulation develops oscillations with frequency $\omega_b$ at an amplitude that decreases with increasing friction coefficient $\gamma$. Assuming that cilia are immersed in water with viscosity, $\eta = 8.5 \times 10^{-4}\mathrm{pN\cdot s/\mu m^2}$, we use values $\gamma= 0.1$\,pN\,s/$\mu$m and  $\gamma= 0.3$\,pN\,s/$\mu$m corresponding to experimentally plausible values of bead radii $r= 6.3\mu m$ and $r= 18.8 \mu m$ respectively under the assumption of Stokes' law $\gamma=6\pi\eta r$~\cite{kotar2010hydrodynamic,bruot2016realizing}. For these friction values, we estimate excess heat fluxes' amplitudes to be of the order of tenths of zeptowatts ($\sim 10^{-22}$W). Such order of magnitude is currently inaccessible  by direct heat-flux measurements by the most accurate $\sim$nW micro-calorimeters, yet measurable indirectly through Eq.~\eqref{powerrowing} via   standard, nanometer-resolution  tip position fluctuations.    

\begin{figure}[h]
 \centering
      \begin{subfigure}{0.49\textwidth}
         \centering
         \def\svgwidth{0.8\linewidth}        
        \includegraphics[scale = 0.9]{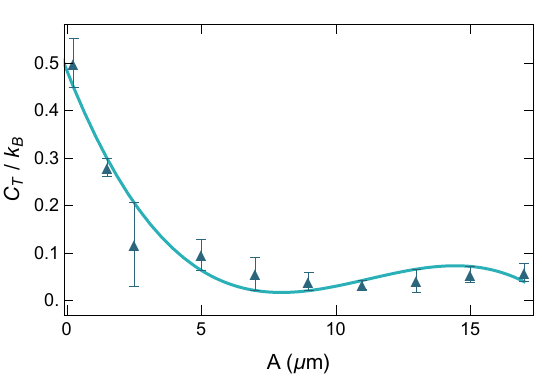}
        \caption{} \label{hcgamma-a}
     \end{subfigure}
     \hfill
     \centering
      \begin{subfigure}{0.49\textwidth}
         \centering
         \def\svgwidth{0.8\linewidth}        
        \includegraphics[scale = 0.9]{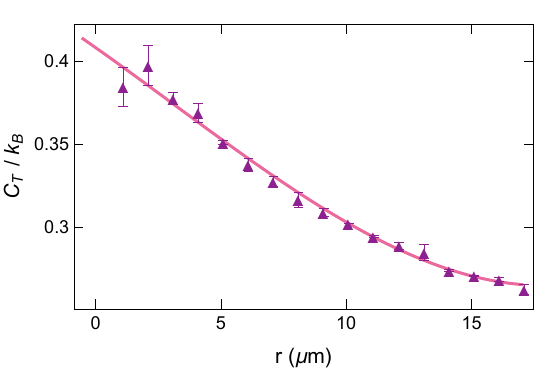}
        \caption{}
        \label{hcgamma-b}
     \end{subfigure}
\caption{{\bf Nonequilibrium heat capacity in the rower model as a function of biophysical parameters.} Heat capacity at temperature $T=300K$  using AC-calorimetry with $T_b(t) = T \,(1 - \epsilon_b \sin(\omega_b t))$, with $\epsilon_b=0.01$ and $\omega_b=2\pi$Hz:  (a) as a function of $A$, half the distance between the minima of the potentials with  $r=6.25 \mu m $ (or equivalently $\gamma =0.1\,\mathrm{pN\cdot s}/\mu\mathrm{m}$), and (b) as a function of $r$, the bead radius of the tip for  $A = 0.5\,\mu\mathrm{m}$.   The symbols are results from numerical simulations and the solid lines fits to empirical functions.  The simulation is performed for $10^4$ trajectories over $200\,\mathrm{s}$, corresponding to an average over $200$ oscillation periods, with $a = 0.25\,\mu\mathrm{m}$, $\kappa = 1.5\,\mathrm{pN}/\mu\mathrm{m}$.   The expected heat flux is  obtained from \eqref{powerrowing} and the heat capacity  using Eqs.~\eqref{mama} and~\eqref{cac}. Error bars are given by standard error of the mean. The fits are: (a) $C_T(A)/k_B = d_0 + d_1 A + d_2 A^2 + d_3 A^3$, with  $d_0 = 0.49$, $d_1 = -0.14\,\mu\mathrm{m}^{-1}$, $d_2 = 0.013\,\mu\mathrm{m}^{-2}$, and $d_3 = -4.0 \times 10^{-4}\,\mu\mathrm{m}^{-3}$; and $C_T(r)/k_B = c_0 + c_1 r + c_2 r^2 + c_3 r^3$, (b) with $c_0 = 0.41$, $c_1 = -1.1 \times 10^{-2}\,\mu\mathrm{m}^{-1}$, $c_2 = 1.6 \times 10^{-4}\,\mu\mathrm{m}^{-2}$, and $c_3 = 1.7 \times 10^{-5}\,\mu\mathrm{m}^{-3}$. }  
\end{figure}

Extracting the first three components of the Fourier-sum decomposition of $\dot{\cal Q}(t)-\dot{Q}_T$ vs $t$ following Eq.~\eqref{mama}  we extract (from the out-of-phase Fourier mode) values of the nonequilibrium heat capacity $C_T$ at fixed $T$, $a$ and $\kappa$ and explore its dependency with $A$ for fixed $\gamma$ (Fig.~\ref{hcgamma-a}) and its dependency on $\gamma$ (through $r=\gamma/6\pi\eta$) for fixed $A$ (Fig.~\ref{hcgamma-b}).  For both scenarios we find numerically that $C_T\geq 0$ is positive for all the explored parameter ranges. At fixed friction coefficient (Fig.~\ref{hcgamma-a}),  the heat capacity decreases with $A$ tending to zero for large $A$ when the heat absorbed at $\pm a$ becomes very small. On the other hand, in the limit $A\to 0$ we get $C_T\to k_B/2$ as the two potentials overlap, retrieving the equilibrium limit. At fixed $A>0$  (Fig.~\ref{hcgamma-b}), $C_T$ also decreases with increasing $\gamma$ towards zero in the very large $\gamma$ limit as expected in the highly damped regime where Brownian fluctuations dominate (and the heat flux vanishes). When $\gamma\to 0$ for fixed $A$, we approach the deterministic limit for which we get $C_T \simeq 0.4 k_B$ for the parameters used in our simulations. We have also explored the dependency of $C_T$ on the temperature  finding a very low impact on the heat capacity values by changing $T$ in the range accessible experimentally between  $300-400$K; see Appendix~\ref{sec:moreon}. 

\newpage
\subsection{Nonequilibrium calorimetry of  ciliar beating II: Markov jump  model}
\label{dis}

In the preceding subsection we have studied a continuous model for ciliar beating which assumed that switches (energy pumps) take place swiftly as soon as a first-passage criterion is met. In a more realistic scenario, such energy pump may be triggered or not with a certain probability when a criterion is first met, and after a stochastic waiting time. To include such realistic ingredients, we build and analyze a continuous-time Markov jump model for the ciliar beating  which includes a finite rate of energy pumping when the cilium is in a specific configuration. The reason for taking a Markov-jump description is also to explore analytical avenues for the heat capacity using the quasipotential method instead of AC-calorimetry.

\begin{figure}[H]
\centering
  \begin{subfigure}{0.46\textwidth}
    \centering
    \includegraphics[scale=0.7]{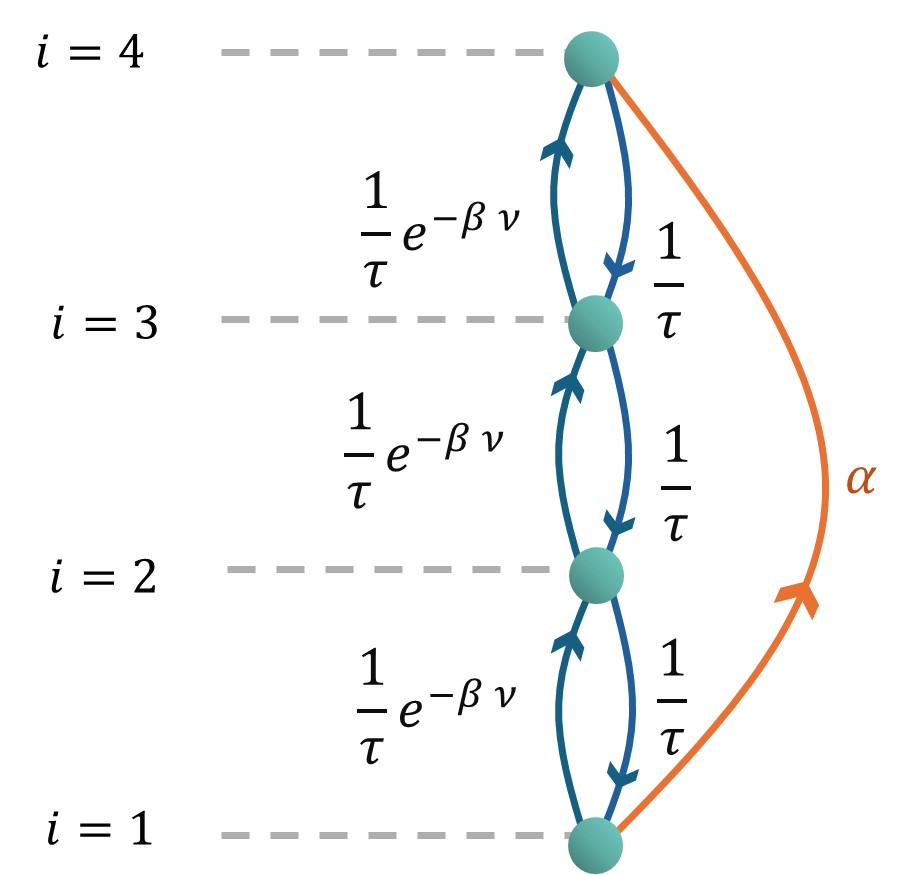}
    \caption{}  \label{rowjump-a}
  \end{subfigure}
  \begin{subfigure}{0.46\textwidth}
    \centering
    \includegraphics[scale=0.9]{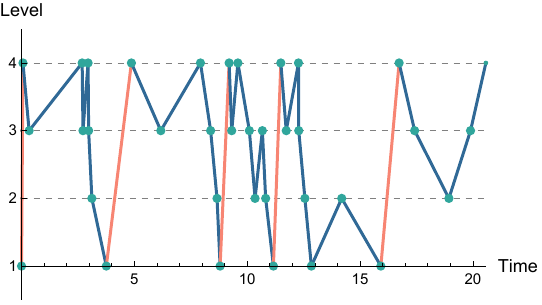}
       \caption{}  \label{rowjump-b}
  \end{subfigure}
  \caption{{\bf Continuous-time Markov jump version of the cilium rower model.} \small{(a) Sketch of a four-level discrete rower model with states (levels, green circles) labeled by the index $i$, and  rates  of the allowed transitions indicated close to their corresponding lines with arrows. (b) Example   trajectory obtained from a numerical simulation for the $4-$state rower model for values $\beta=1/k_B T$, $\tau=1$s (intra-level characteristic time), $\alpha=0.5$Hz (rate of energy injection) and $\nu=2 k_B T $ (energy barrier amplitude).}} 
\end{figure}
The model, sketched in Fig.~\ref{rowjump-a} is as a follows: the cilium is described by a continuous-time evolution in a discrete energy space (on an $n$-level graph), where the energies of the levels range from $E_1=0$ to $E_n=(n-1)\nu$, mimicking a harmonic oscillator.  Let $\beta=1/ k_B T$, where $T$ is the temperature of the heat bath to which the rower is weakly coupled, and $k_B$ is the Boltzmann constant.
Thermal transitions between adjacent energy levels are allowed: for jumping  up, $n\to n+1$, the transition rate is given by $e^{-\beta \nu}/\tau$, while for jumping down, $n+1\to n$, the transition rate is $1/\tau$.   Furthermore, in analogy with  the energy switch in the continuous model, an energy-pumping mechanism is introduced with an energy input $E_n=(n-1)\nu$ to the rower from the lowest-energy level $E_1=0$. In particular, we model such a switch as unidirectional with rate $\alpha>0$ from $1\to n$ while forbidden from $n\to 1$ whenever we consider $(n>2)$ energy levels\footnote{The transition $n\to 1$ is allowed only for $n=2$ (two-level rower) but at the rate of the thermal transition $1/\tau$.}. 
For $\alpha=0$, all transitions are thermal and obey detailed balance, yielding equilibrium dynamics, while for $\alpha>0$ we have a nonequilibrium dynamics where the  nonequilibrium parameters are $\alpha$ (to be compared with $1/\tau$) and the size $n$ (proportional to the pumped energy $E_n=(n-1)\nu$). The dynamics can be made to deviate significantly from equilibrium for large $n$ and/or $\nu$ at $\alpha>0$. 
The fact that the input of energy occurs at a finite $\alpha$ adds a new ingredient compared to the diffusive rower model where the switches are triggered deterministically. In this discrete version, the energy pump starts at a random time whenever the system is at the lowest energy level, with an exponentially distributed waiting time at rate $\alpha$. See Fig.~\ref{rowjump-b} for an example numerical simulation of the model.

\begin{figure}[H]
 \centering
      \begin{subfigure}{0.49\textwidth}
         \centering
         \def\svgwidth{0.8\linewidth}        
        \includegraphics[scale = 0.4]{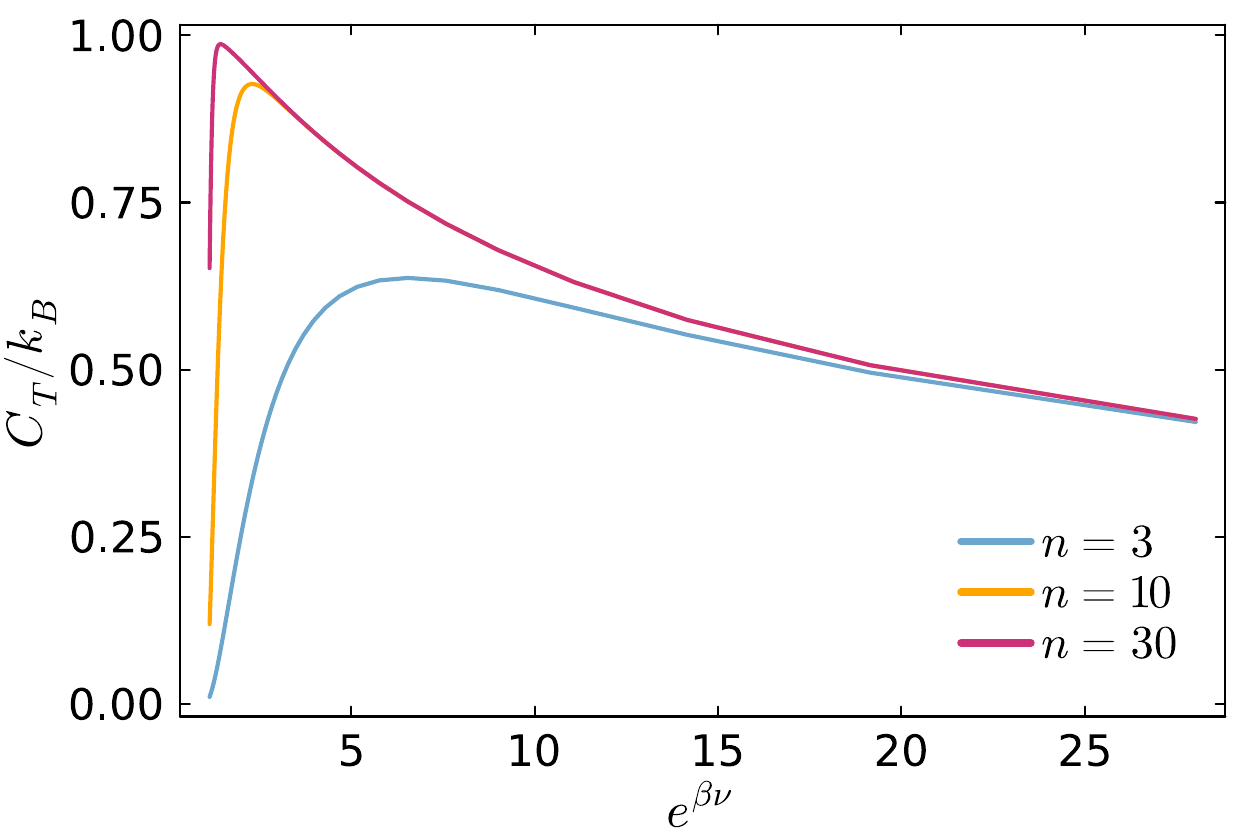}
        \caption{}  \label{Ceqrowermarkova}
     \end{subfigure}
     \hfill
     \centering
      \begin{subfigure}{0.49\textwidth}
         \centering
         \def\svgwidth{0.8\linewidth}        
        \includegraphics[scale = 0.4]{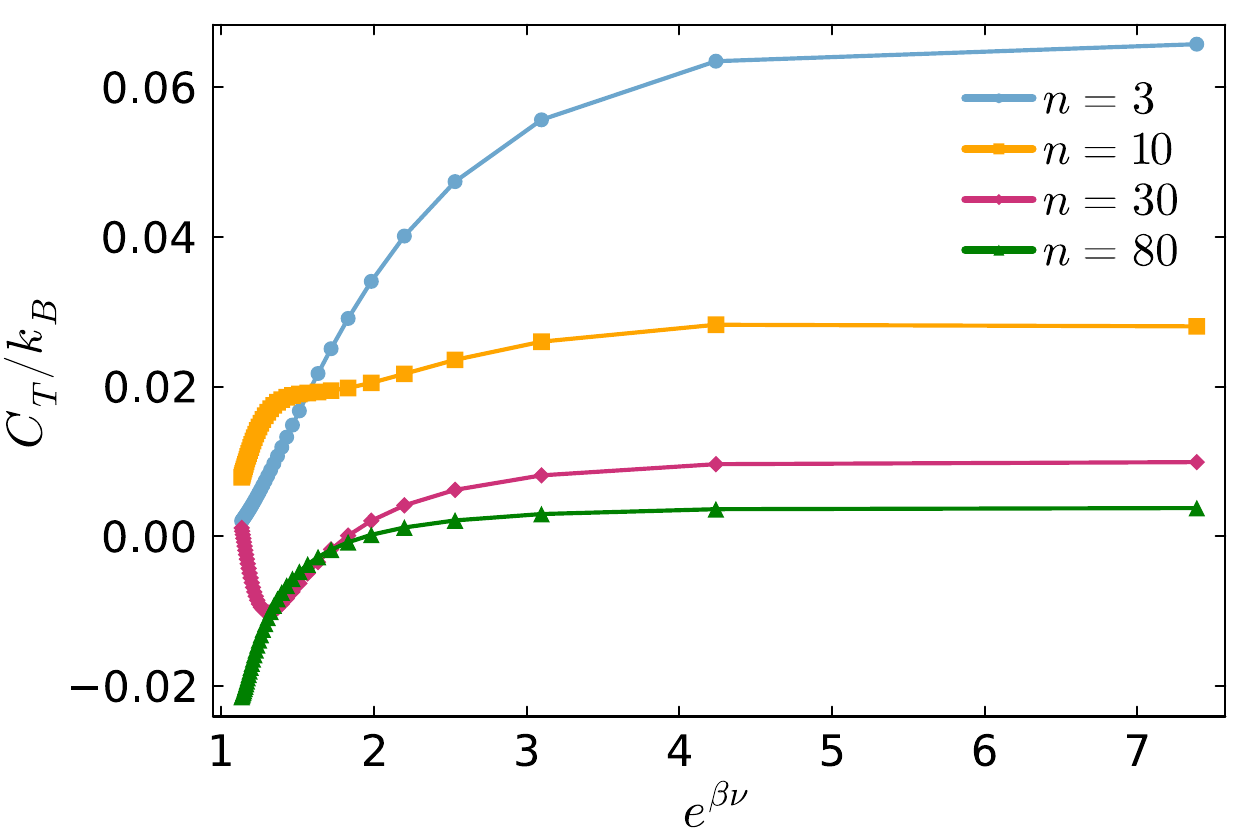}
        \caption{}
        \label{Ceqrowermarkovb}
     \end{subfigure}
\caption{{\bf Heat capacity in equilibrium (a) and in nonequilibrium (b) conditions as a function of the energy-level spacing $\nu$ (in $k_B T $ units) for the discrete rower model for ciliar beating sketched in Fig.~\ref{rowjump-a}.  }\small{Heat capacity associated with the Markov-jump model for cilium motion sketched in Fig.~\ref{rowjump-a} as a function of $e^{\beta\nu}$, with~$\nu$ the energy-level spacing and $\beta=1/k_B T$, for different values of the number of levels $n$ (see legends): (a) in equilibrium $\alpha=0$, and in (b) nonequilibrium conditions $\alpha=1$. The lines in (a) are obtained from the analytical expression Eq.~\eqref{rowE}, while the symbols connected by lines in (b) are obtained via the quasipotential method described in Appendix~\ref{M1}. } }
\end{figure}

We first focus on the  equilibrium limit  $\alpha=0$, where we can use the canonical fluctuation formula
\begin{equation}
C_T^{\text{eq}}=\partial_T\langle U\rangle^{\text{eq}}=\mathrm{Var}(U)/(k_BT^2),
\label{eq:canonical}
\end{equation}
where the variance of the energy $\mathrm{Var}(U)$ can be easily retrieved using the canonical ensemble. That equilibrium heat capacity is independent of $\tau$ and from a straightforward calculation using Eq.~\eqref{eq:canonical}, we find 
\begin{equation}\label{rowE}
C_T^{\text{eq}}=k_B \, \beta ^2 \nu^2  \left[\frac{ e^{\beta  \nu }}{\left(e^{\beta  \nu }-1\right)^2}-\frac{n^2  e^{n \beta  \nu }}{\left(e^{n \beta  \nu }-1\right)^2}\right].
\end{equation}
Notably, this analytical result reveals that the equilibrium heat capacity of this model depends on the energy-level spacing $\nu$ in a non-monotonous way for all values of $n$, and we retrieve the Third Law of thermodynamics with vanishing heat capacity at absolute zero or as $\nu=0$. Increasing $\nu>0$, the heat capacity increases at small $\nu$ till it reaches a local maximum --- the so-called Schottky peak --- after which it decreases monotonously towards zero for large energy spacing (see Fig.~\ref{Ceqrowermarkova}). Our choice to plot  $C_T^{\text{eq}}$ in Fig.~\ref{Ceqrowermarkova} as a function of  $e^{\beta\nu}$ is to analyze the dependence on the energy spacing in $k_{B} T $ units, but an analogous effect is found when varying the temperature, finding a local maximum of the heat capacity as a function of the temperature (Schottky anomaly). 
Furthermore, analyzing how $C_T^{\text{eq}}$  vs $e^{\beta\nu}$ depends on the number $n$ of energy levels, we find from Fig.~\ref{Ceqrowermarkova}  that the equilibrium heat capacity is bounded between zero (equilibrium limit) and Boltzmann's constant, i.e.,
\begin{equation}
0\leq C_T^{\text{eq}} \leq k_{B}.\label{cteqrowerdic}
\end{equation}
Notably, the upper limit $k_{B}$ is approached asymptotically for a large number of levels $n$ in the $\beta\nu\to 0$ limit (small energy spacing and/or at high temperature). 
We evaluate the nonequilibrium heat capacity $C_T$ associated with the rower Markov-jump model following the steps outlined in Appendix~\ref{M1}, that is,  using \eqref{hca} after solving the Poisson equation \eqref{quasi} to obtain the quasipotential, see also Appendix~\ref{rowmore} for additional details specific to this model. 
We first focus in  Fig.~\ref{Ceqrowermarkovb} on the dependence of $C_T$ {\it vs} $e^{\beta\nu}$ for a pump rate $\alpha$ (in units of the thermal waiting time $\tau$). For small $n$, $C_T$ is positive and has a similar phenomenology to the equilibrium scenario with a local maximum at intermediate temperatures.  Intriguingly, for $n$ large, $C_T$ can take negative values (inaccessible in the equilibrium scenario by virtue of Eq.~\eqref{cteqrowerdic}) and even display an inverted Schottky anomaly ---a genuine nonequilibrium feature--- with a local minimum as shown for the $n=30$ case in Fig.~\ref{Ceqrowermarkovb}.   


\begin{figure}[H]
 \centering
      \begin{subfigure}{0.49\textwidth}
         \centering
         \def\svgwidth{\linewidth}        
        \includegraphics[scale =  0.38]{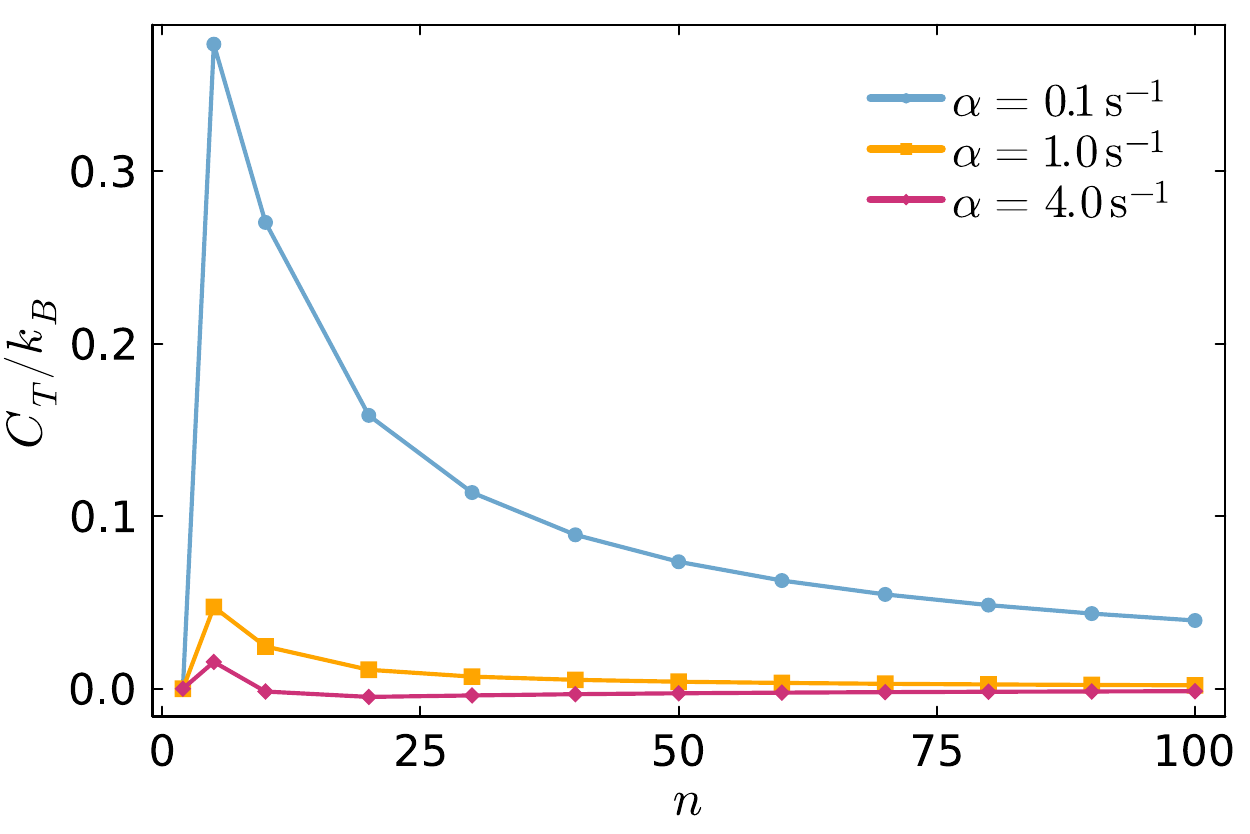}
        \caption{}  \label{CforTa}
     \end{subfigure}
     \hfill
     \centering
      \begin{subfigure}{0.49\textwidth}
         \centering
         \def\svgwidth{0.8\linewidth}        
        \includegraphics[scale = 0.38]{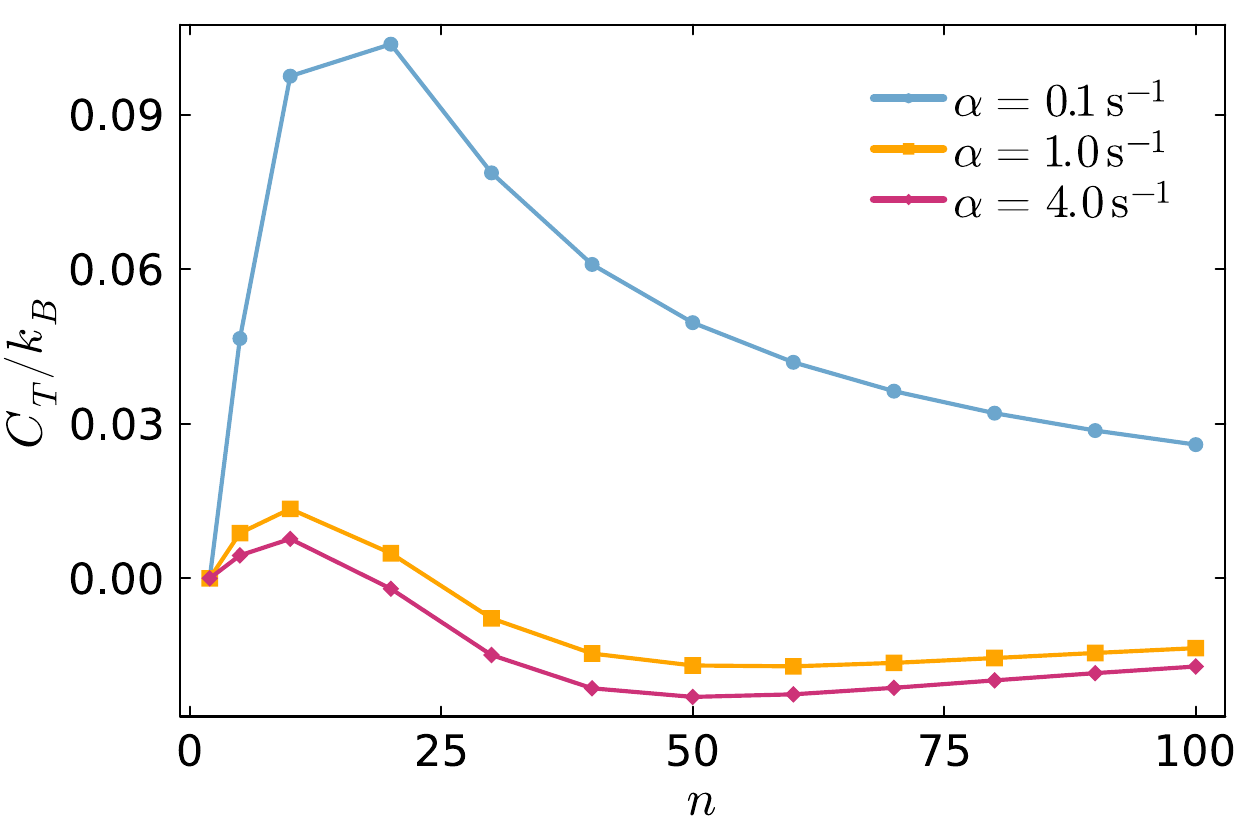}
        \caption{}
        \label{CforTb}
     \end{subfigure}
\caption{{\bf Heat capacity as a function of the number of energy levels  $n$ and the energy injection rate  for the discrete rower model for ciliar beating sketched in Fig.~\ref{rowjump-a} at a lower (a) and higher (b) temperature values.  }\small{The heat capacity of the Markov-jump model for cilium motion sketched in Fig.~\ref{rowjump-a} as a function of the number of levels $n$ for different values of the energy injection rate $\alpha$ (see legends): (a) at temperature  $T=1$, and at temperature (b) $T=5$. The symbols  are obtained via the quasipotential method described in Appendix~\ref{M1}, and lines are a guide to the eye. In both panels, we set the energy spacing $\nu=k_{B}T$ and the thermal waiting time $\tau=1$.} }
\end{figure}

To further explore the negativity of the nonequilibrium heat capacity at large number $n$ of energy levels, we consider  $C_T$  as a function of $n$  for different energy injection rates $\alpha$ at two temperature values $T=1$ (Fig.~\ref{CforTa}) and $T=5$ (Fig.~\ref{CforTb}).  At the lower temperature $T=1$, the nonequilibrium heat capacity decreases its value, taking only positive values, when increasing $\alpha$ in the range that we analyzed, with a similar phenomenology as in the standard Schottky anomaly with a local maximum in this case at moderate $n$.  On the other hand, at the higher temperature value $T=5$ we find that by increasing $\alpha$, $C_T$ reaches negative values developing even two local extrema as a function of $n$ with a local positive maximum at $n\simeq 10$ followed by a local negative minimum at $n\simeq 50$ for the considered parameters.

Taken together, the results of our analyses for the continuous and discrete `rower' models for the ciliar motion illustrate that the  heat capacity depends non-monotonously on key parameters of the model and can take both positive and negative values with the order of magnitude of its absolute value of around $k_B$ or less.  In the next Sec.~\ref{sec:motor}, we explore nonequilibrium calorimetry in a minimal model for molecular motor motion to seek whether or not the $\sim k_B$ order of magnitude and the negativity of $C_T$ also extends to a broader biophysical scenario.  The occurrence of a negative heat capacity (for systems that are weakly coupled to a heat bath) is only possible for nonequilibrium steady states, and signifies a separation between heat (and associated Clausius etropy) and occupation (and associated Boltzmann entropy), as explained in \cite{negheat}.



\section{Heat capacity of molecular motor motion}
\label{sec:motor}
Molecular motors are nanoscale proteins that convert chemical free energy, most often from  hydrolysis of adenosine triphosphate (ATP), into  mechanical motion along  tracks. Such directed motion is essential for fast and reliable intracellular transport of e.g. vesicles against  drag forces.   Their operation is inherently stochastic, driven by the interplay between potential forces, thermal fluctuations, and nonequilibrium chemical reactions~\cite{Ajdari,keller2000mechanochemistry,howard2002mechanics,kolomeisky2007molecular,Fulga2009, Astumian2010, Reimann2002}. 

Although molecular motor motion is considered a paradigm of nonequilibrium physics in the biological realm, most of the studies regarding their thermodynamics   have focused so far on their steady-state properties such as mean velocity and power, rate of heat dissipation,  average efficiency, and current fluctuations among others~\cite{howard2002mechanics,pietzonka2016universal,baiesi2018life,guillet2020extreme,li2020efficiencies}. Here, as a first step to tackle motors' nonequilibrium response to thermal perturbations,  we focus on analyzing the dependency of the  heat capacity of molecular motors as a function of parameters controlling their activity. To tackle this quest, we focus on popular theoretical models by working with  both continuous (Langevin) and discrete-state (Markov jump) dynamics given the broad applicability of both approaches in molecular motor biophysics.

\subsection{Nonequilibrium calorimetry of a molecular motor:  Langevin  (diffusion)  model}
A minimal yet powerful way to describe such systems is through a hybrid model combining a continuous diffusive coordinate, representing the motor’s  center-of-mass position along a track, with discrete internal states corresponding to different  configurations (e.g. bound/unbound); see {\it e.g.},  the Review~\cite{Ajdari} which we use as a reference modeling framework. There we assume only two configurations, one corresponding to the motor being attached to a track and another corresponding to the motor being detached, making a two-state diffusive-coordinate framework with binding/unbinding cycles, force generation, and directional bias induced by the interplay between ATP consumption, external forces, and potential broken  (a)symmetries. 

\begin{figure}[h]
  \centering
  \begin{subfigure}[b]{0.49\textwidth}
    \centering
     \includegraphics[scale=0.38]{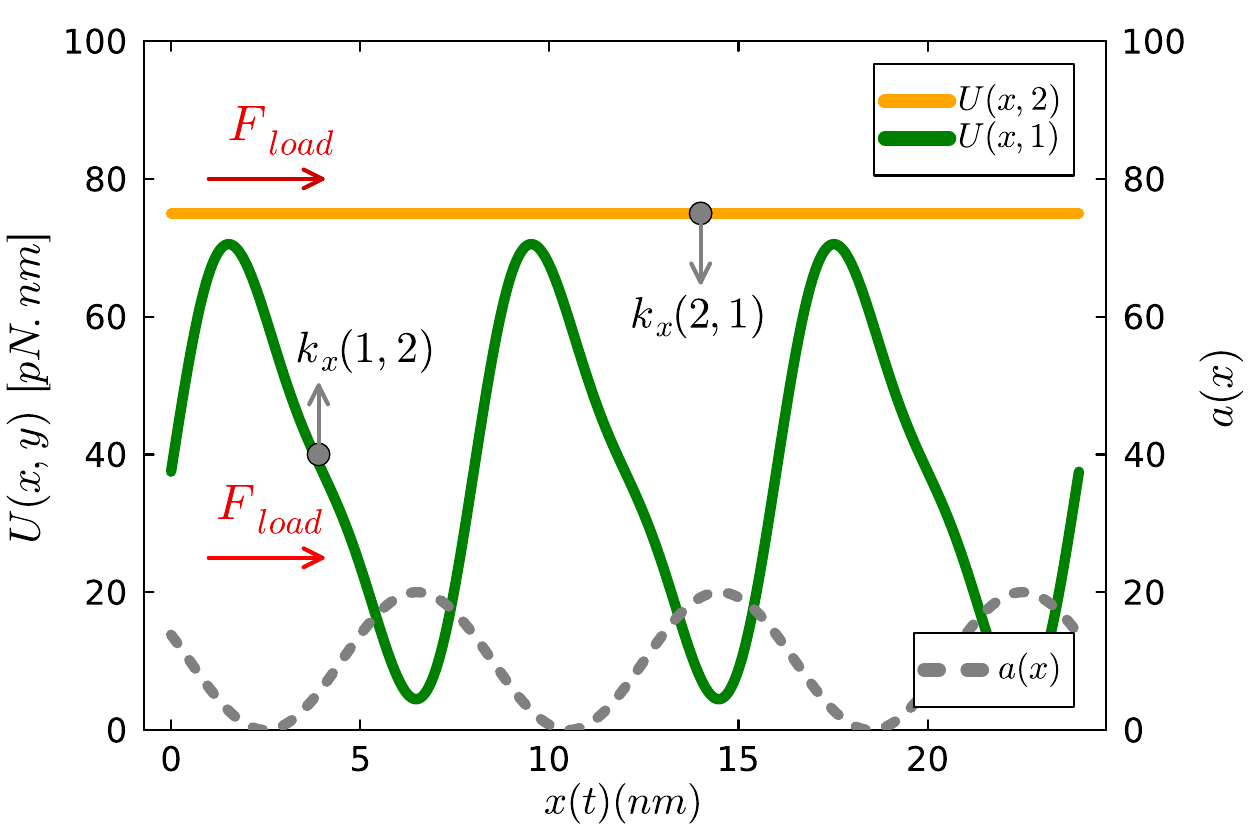}
    \caption{}
     \label{fig:sub-filament}
      \end{subfigure}
  \hfill
  \begin{subfigure}[b]{0.49\textwidth}
    \centering
     \includegraphics[scale = 0.375]{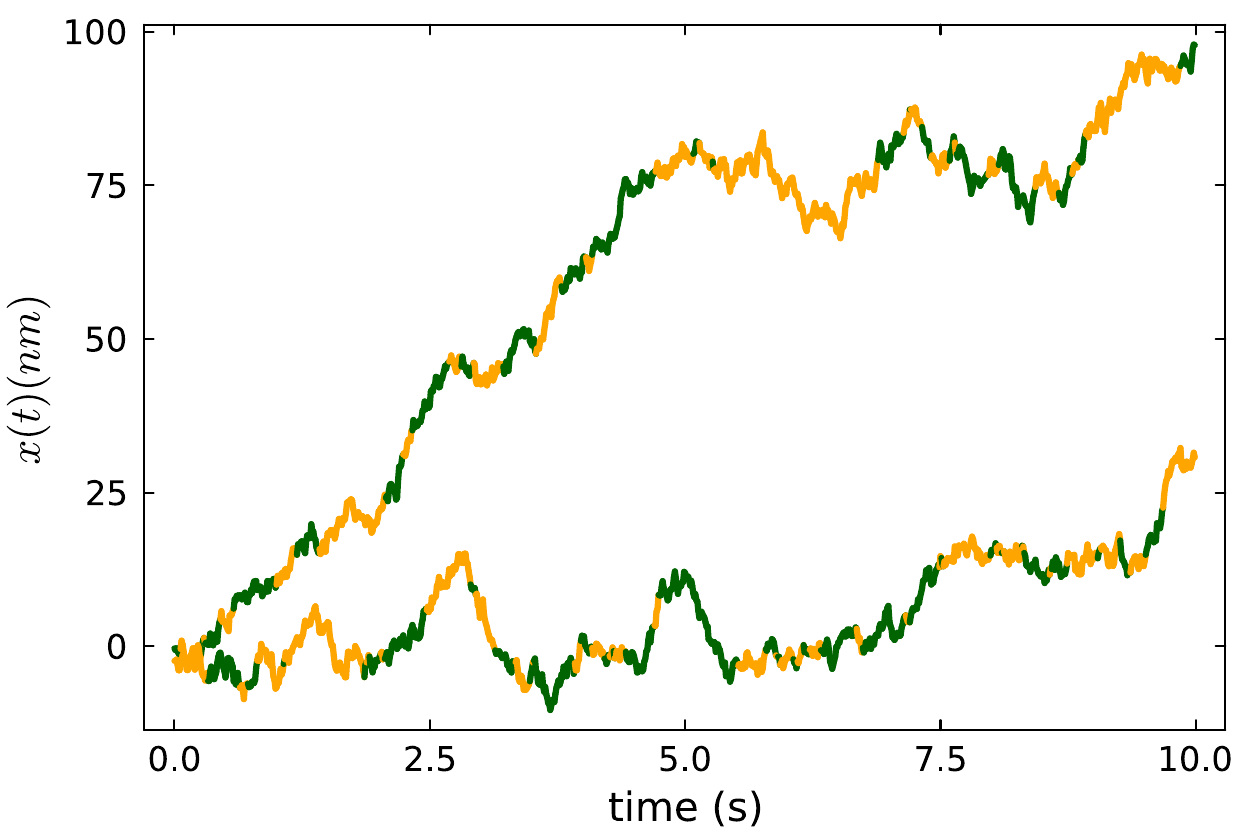}
    \caption{}
    \label{fig:sub-mm}
  \end{subfigure}
  \caption{{\bf Sketch of continuous flashing ratchet model for molecular motor motion and example trajectories from numerical simulations. }\small{(a) Sketch of the model given by Eqs.~(\ref{2lajdari}-\ref{eq:activesites}).   The motor follows overdamped diffusive dynamics in a potential that switches between two states, $y=1$ and $y=2$, and in each state. An external force $F_{\text{load}}$ acts in both states.  The potential (green line) is  skewed periodic for $y=1$ [Eq.~\eqref{eqmot1}] and  flat for $y=2$ [Eq.~\eqref{eqmot2}].  The skewed potential is plotted along $x$ for values $u_1= 30\,$ pNnm, asymmetry parameter $\sigma=0.25$, and spatial period  $\lambda= 8$ nm. The flat potential is constant, $U(x,2)=75$ pNnm. The dashed gray line is an example space-dependent rate $a(x)$ for chemically-induced transitions in the `active sites' scenario [Eq.~\eqref{eq:activesites}] with maxima near the minima of the potential. (b) Example time series obtained for parameter values  $u_1 = 3 k_{B}T$, $u_2 = 7.5   k_{B}T$, $\Delta \mu =k_{B}T$, $a_0=2$Hz, and $c_0=1$Hz. The yellow and green shaded regions correspond to the time instances where the potential is $U(x,2)$ and $U(x,1)$ respectively.}}
  \label{filapot}
\end{figure}
The molecular motor dynamics couoples an overdamped Langevin dynamics in asymmetric periodic potentials with a two-level jump process.
The motor protein is modelled with two conformational states, $y(t) \in \{1,2\}$.  In the bounded state ($y=1$), the motor interacts with a periodic, asymmetric potential along $x$  that reflects the motion on a polar periodic track. 
In the unbounded state ($y=2$), the motor diffuses freely on a flat potential surface. The motor’s position along the track  $x(t)$  follows    the Langevin equation
\begin{equation}\label{2lajdari}
    \gamma \dot{x}(t) = - \, \partial_x U(x(t), y(t))+ F_{\text{load}}+ \sqrt{2 k_B T \gamma} \, \xi(t),
\end{equation}
where $\gamma$ is the friction coefficient,  $U(x,y)$  is a space-dependent  potential for both the bounded ($y=1$) and unbounded ($y=2$) configurations, both $U(x,y=1)$ and $U(x,y=2)$ being periodic in space and defined below.  Eq.~\eqref{2lajdari} also considers the presence of a constant external load force $F_{\text{load}}$ (positive in the direction of forward motion) that is exerted on the motor irrespective of whether it is in a bonded or unbounded state. The last term in \eqref{2lajdari} accounts for thermal fluctuations at temperature $T$, with $\xi$ a unit variance Gaussian white noise with  $\langle \xi(t) \rangle = 0$ for all $t$, $\langle \xi(t) \xi(t') \rangle = \delta(t - t')$. For the potential, we use 
\begin{eqnarray}
U(x,1)&=&u_1\left[\sin\left(\frac{2 \pi x}{\lambda}\right)+\sigma \sin\left(\frac{4 \pi x}{\lambda}\right)+(1+\sigma)\right],\label{eqmot1}\\
U(x,2)&=&u_2.\label{eqmot2}
\end{eqnarray}
For $y=1$ the motor moves in the periodic asymmetric potential given by Eq.~\eqref{eqmot1}, while for   $y=2$ the potential is  $u_2$ independent of $x$  [Eq.~\eqref{eqmot2}].
 Here,  the parameters $\sigma$  controls the difference in up/downward slopes, 
and $\lambda$ is the spatial period of the potential.  See Fig.~\ref{fig:sub-mm} for an illustration of the potentials for specific parameter values. 
 
\begin{table}[h]
\centering
\caption{{Parameters used in the flashing-ratchet model for molecular motor motion \eqref{2lajdari}--\eqref{random}.}}\label{tableparam}
\scriptsize
\begin{tabular}{|c|c|c|p{10cm}|}
\hline
{\bf Parameter} & {\bf Value} & {\bf Dimensions} & {\bf Explanation} \\
\hline
$\gamma$   & $ 0.1$   & pN$\cdot$s/nm & Effective 1D drag coefficient of the motor along the filament \\
\hline
$\lambda$     & $8$   & nm  & Periodicity of the track \\
\hline
$u_{1}$    & $30$  & pN$\cdot$nm   & Amplitude of bound-state potential \\
\hline
$u_{2}$    & $75$   & pN$\cdot$nm   & Reference energy for the flat potential for the unbound/diffusive state  \\
\hline
$\sigma$   & $0.25$  & --            & Asymmetry parameter of the periodic potential\\
\hline
$k_{B}$    & $0.0138$ & pN$\cdot$nm/K & Boltzmann constant in  dimensions relevant to molecular motor motion \\
\hline
$a_0$    & $20$  & s$^{-1}$      & Baseline chemically-driven transition rate\\
\hline
$c_0$    & $1$   & s$^{-1}$      & Baseline thermally-activated transition rate \\
\hline
$\Delta G$ & $100$ & pN$\cdot$nm & ATP hydrolysis free energy at room temperature ($\sim 25 k_{B}T$)\\
\hline
\end{tabular}
\end{table}

The  dynamics of $x_t$ is coupled to the instantaneous value $y_t$ of the internal state. The internal state $y_t$   follows a continuous-time  two-state  Markov dynamics controlled by the coupling between chemical and mechanical interactions. Such coupling is modelled via an explicit  space-dependence of the rates $k_x(1,2)$ [$k_x(2,1)$] for the transitions $y=1 \to y=2$  [$y=2\to y=1$].  Inspired by Refs.~\cite{Ajdari,J_licher}, we use 
\begin{align}\label{random}
    k_x(1,2) = [c_0+ a(x)e^{{\beta}\Delta G} ]\, e^{  \beta \, U(x,1) }, \qquad  k_x(2,1)= [c_0+a(x)] \, e^{ \beta \, U(x,2) }.
\end{align}
 given that the motor position is $x$.
 Here, $c_0$ is a (space-independent) characteristic  rate constant of thermal transitions, and $\Delta G = G_{\text{ATP}} - G_{\text{ADP}} - G_{P}$ is the Gibbs free energy change associated with the chemical reaction $\text{ATP} \rightarrow \text{ADP} + \text{P}$ of hydrolysis of one adenosine triphosphate (ATP) molecule into adenosine diphosphate (ADP) and inorganic phosphate (P). We chose
 \begin{equation}\label{eq:activesites}
 a(x) =a_0 \cos^2\left[\frac{\pi}{\lambda}\left(x - \frac{3}{4}\lambda - \frac{1}{2}\right)\right],
 \end{equation}
  to describe the `active sites' paradigm in which the chemical fuel consumption is localized near the potential minima, as suggested by a plethora of experimental and theoretical studies in molecular motor motion~\cite{howard2002mechanics, Ajdari}. For simplicity, we take here the same $\lambda$ as appears in \eqref{eqmot1} for the potential $U(x,1)$, indicating the same periodicity.

\begin{figure}[H]
\centering
      \begin{subfigure}{0.49\textwidth}
         \centering
         \def\svgwidth{0.8\linewidth}        
        \includegraphics[scale = 0.8]{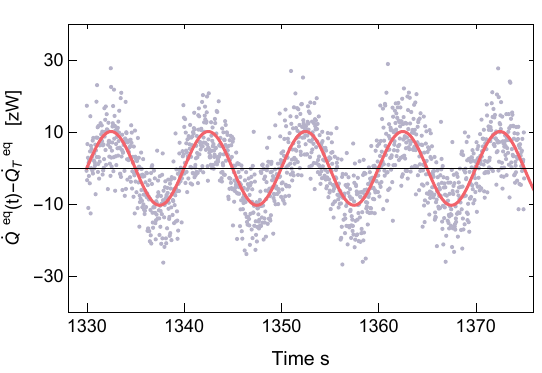}
        \caption{\small{}}
        \label{powerRatcheta}
     \end{subfigure}
      \hfill
     \centering
      \begin{subfigure}{0.49\textwidth}
         \centering
         \def\svgwidth{0.8\linewidth}        
        \includegraphics[scale = 0.8]{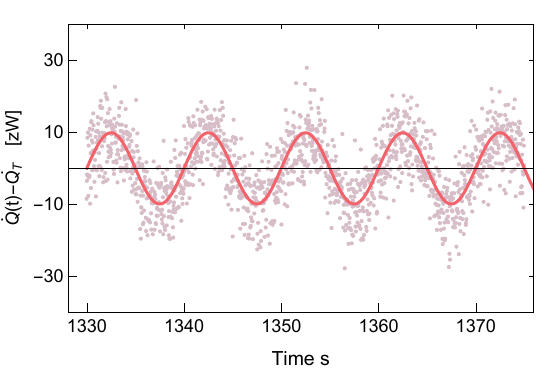}
            \caption{\small{}}
            \label{powerRatchetb}
     \end{subfigure}
\caption{{\bf{Expected excess heat flux as a function of time for the continuous flashing ratchet model for molecular motor motion obtained from AC calorimetry (a) in equilibrium, and (b)  out of equilibrium.}} 
\small{ Heat flux of the dynamics in \eqref{2lajdari}, for the parameter values in Table~\ref{tableparam}, with $\Delta G=0$,   $T = 290$ K,  $\varepsilon_b = 0.05$, $\omega_b = (\pi/5)$Hz, and load force values $F_{\text{load}}= 0 pN$ (a) and $F_{\text{load}}= 2 pN$ (b). We show only a small time window of the heat flux for the sake of visualization,  obtained by averaging over $10^5$ samples, and the simulation is performed for a duration of $2000\,\mathrm{s}$. The scattered dots represent simulation data, and the red solid lines are the fitted curves with the Fourier sum~\eqref{mama}. From the fits, we obtain  (a) the equilibrium value $C_T^{\text{eq}} = 0.62\, k_B$ and (b) $C_T=0.50\, k_B $ for out of equilibrium.}}
\end{figure}

As a benchmark, we first focus on the equilibrium scenario with  $\Delta G = 0$ and $F_{\text{load}} = 0$, for which we evaluate the heat flux  via AC-calorimetry and employing Eq.~\eqref{heateq} to determine the excess heat flux,  see Fig.~\ref{powerRatcheta}. For this model we get excess heat fluxes of the order of zeptowatt (zW).
Then, we extract the first three components of the Fourier-sum decomposition of $\dot{\cal Q}(t)-\dot{Q}_T$ vs $t$ following Eq.~\eqref{mama} to determine the heat capacity $C_T$ which yields $C_T^{\text{eq}} \simeq 0.62\,k_B$ at $T = 290\,\mathrm{K}$. This result is in excellent agreement with the analytical calculation $C_T^{\text{eq}}\simeq 0.61\,k_B$ using  the  canonical fluctuation formula [Eq.~\eqref{eq:canonical}] for the heat capacity,  where $\langle U^n\rangle=\frac{1}{Z}\sum_{y=1}^2\int_0^\lambda U(x,y)^n e^{-\beta U(x,y)}\id x$ with $Z$ the partition function. For the two–state potential $U(x,1)$ and $U(x,2)$, numerical evaluation at $T=290\,\mathrm{K}$ with the given parameters yields $\langle U\rangle\approx 1.64 k_B T$, $\mathrm{Var}(U)\approx 0.62 (k_B T)^2$.

\begin{figure}[H]
\centering
      \begin{subfigure}{0.49\textwidth}
         \centering
         \def\svgwidth{0.8\linewidth}        
        \includegraphics[scale = 0.9]{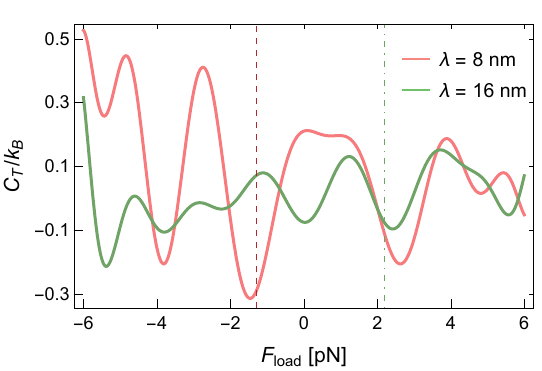}
        \caption{{}} \label{hcr2lm100a}
     \end{subfigure}
      \hfill
     \centering
      \begin{subfigure}{0.49\textwidth}
         \centering
         \def\svgwidth{0.8\linewidth}        
        \includegraphics[scale = 0.9]{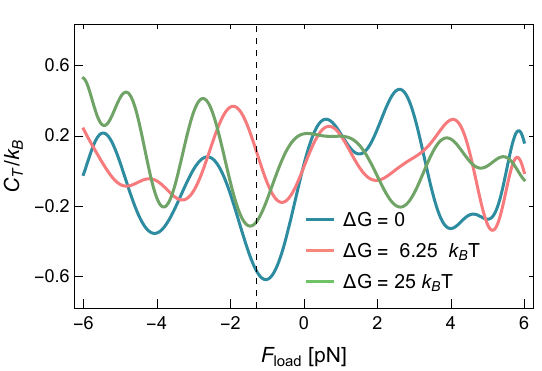}
            \caption{{}} \label{hcr2lm100b}
     \end{subfigure}
\caption{\small {\textbf{Heat capacity of flashing-ratchet model for molecular motors vs external force value  $F_{\text{load}}$}. The vertical non-solid lines represent the stall force at different values.  (a) Different values of $\lambda$ at $\Delta G=25 k_B T$. The dotted red line denotes $F_{\text{stall}}(\lambda=8)=-1.2 $pN for the value of $\lambda =8 $nm
and the green dot--dashed line represent $F_{\text{stall}}(\lambda=16)=2.2 $pN for the value of $\lambda =16 $nm (b). Different values of $\Delta G$ at $\lambda =8 $nm. The dashed vertical line indicates the stalling force, which is $F_{\text{stall}}=-1.2 $pN. The lines are the result of taking average over  $10^5$ samples for $ 2000$ seconds obtained by  AC calorimetry with base temperature $T=290$K, $\epsilon=0.05$, $\omega=(\pi/5)$Hz, with the rest of the parameters as given in Table~\ref{tableparam}. A 10-element moving average followed by a fit to a polynomial of degree 20 is applied to the raw data obtained from AC calorimetry; see Appendix~\ref{app:numericsratchet} for further details.}}  
\end{figure}

In Figs.~\ref{hcr2lm100a} and.~\ref{hcr2lm100b}, we report estimates of the nonequilibrium heat capacity $C_T$ at room temperature $T=290$K  as  functions of the external load force $F_{\text{load}}$  obtained numerically  after applying an $15$-point moving average in the raw heat flux obtained from AC calorimetry (see Appendix~\ref{app:numericsratchet} for further  details on this data analysis). First we focus on a specific value of the chemical free energy $\Delta G = 25 k_B T$, for which $C_T$ displays a  non-monotonic dependency on the load force with multiple maxima and minima that are more noticeable for smaller spatial periods $\lambda$. 
\\ Interestingly, we observe in Figs.~\ref{hcr2lm100a} and Figs~\ref{hcr2lm100b} that $C_T$ often achieves  local minima near the {\em stall} force $F_{\text{stall}}$, that is the force at which the net (time-averaged) velocity of the motor vanishes. In the next Sec.~\ref{jve} we explore further such result by focusing on an analytically-solvable discrete version of the flashing ratchet model for molecular motor motion.


\begin{figure}[H]
\centering 
             \includegraphics[scale = 0.9]{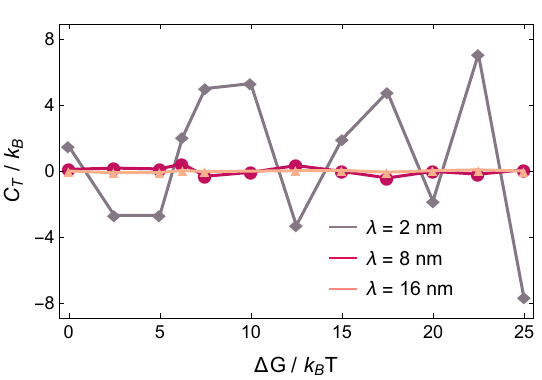}
            \caption{{\bf Heat capacity of flashing-ratchet model for molecular motors: localized vs space-homogeneous ATP hydrolysis. } Nonequilibrium heat capacity $C_T$ for the flashing ratchet sketched in Fig.~\ref{fig:sub-filament} as a function of the ATP hydrolysis free energy for different values of the characteristic length $\lambda$ in the chemical transition rates (see legend). The symbols are obtained using AC calorimetry with base temperature $T=290$K, $\epsilon=0.05$, $\omega=(\pi/5)$Hz  at external force value $F_{\text{load}}=1$pN, with the rest of the parameters as given in Table~\ref{tableparam}. $F_{\text{load}}=1$pN. The lines are a guide to the eye.  We observe a $\sim 10$ fold increase of the heat capacity in going from step sizes of $8$nm to $2$nm}\label{fig:CvsDG}
     \end{figure}

Last but not least, and for the sake of completeness, we  also analyze the dependency of $C_T$ on the chemical free energy $\Delta G$ for different values of the  spatial period $\lambda$,  see Fig.~\ref{fig:CvsDG}. Our numerical estimates reveal a multi-peaked structure for $C_T$  also as a function of $\Delta G$. Notably, visual inspection of the peak-to-peak amplitude of $C_T$ vs $\Delta G$ for various $\lambda$ suggests that motors with smaller step sizes may be more sensitive  to nonequilibrium thermal response. We attribute this result to kinetic effects resulting in a  higher likelihood for excursions above the potential barrier to take place for $\lambda$ small at fixed value of the barrier height $u_1$.  

\subsection{Nonequilibrium calorimetry of a molecular motor:  Markov jump model}\label{jve}

\begin{figure}[H]
   \centering
      \begin{subfigure}{0.6\textwidth}
         \centering
         \def\svgwidth{\linewidth}        
        \includegraphics[scale = 0.85]{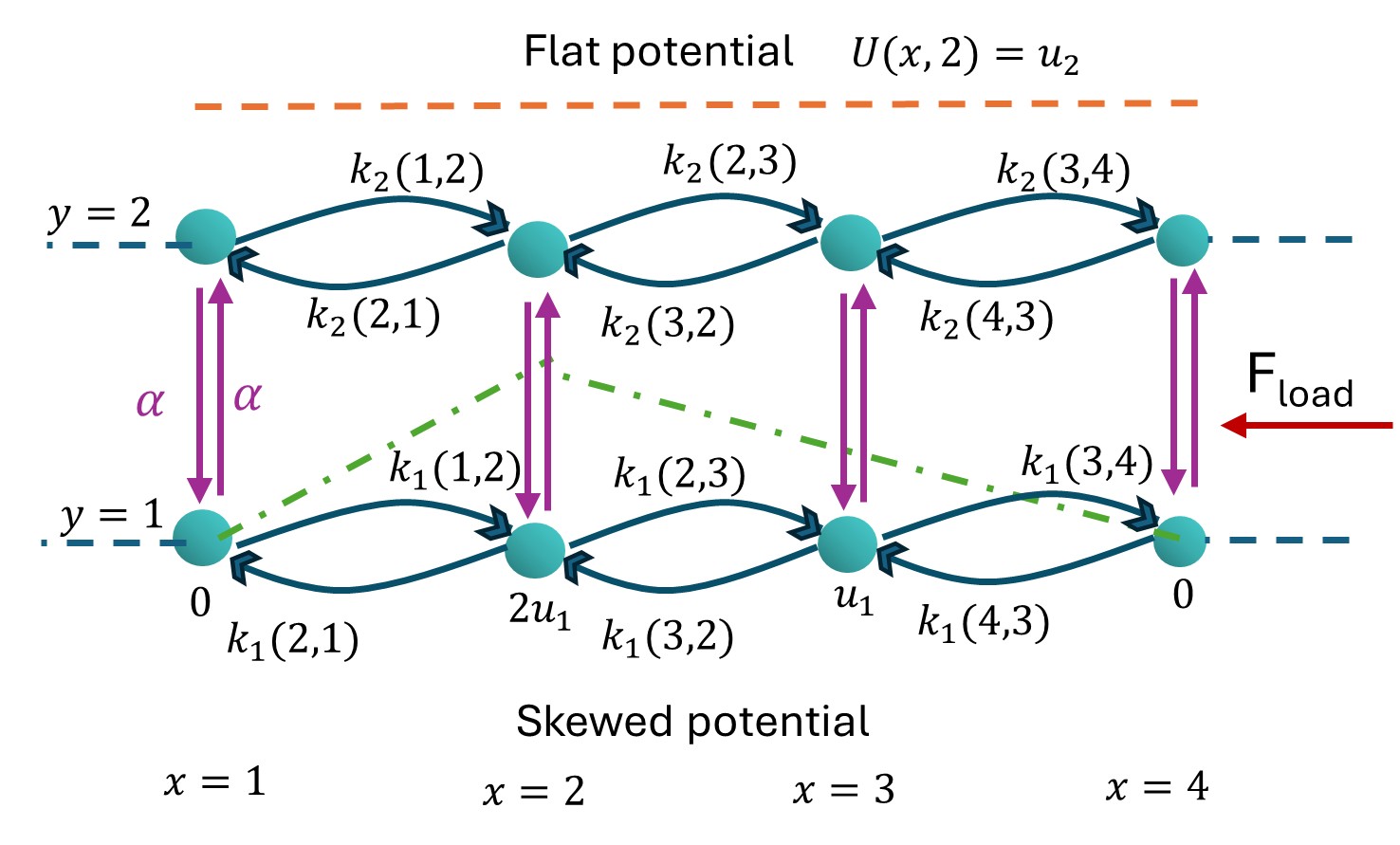}
        \caption{{}}
        \label{dismm3a}
     \end{subfigure}
      \hfill
     \centering
      \begin{subfigure}{0.38\textwidth}
         \centering
         \def\svgwidth{\linewidth}        
        \includegraphics[scale = 0.73]{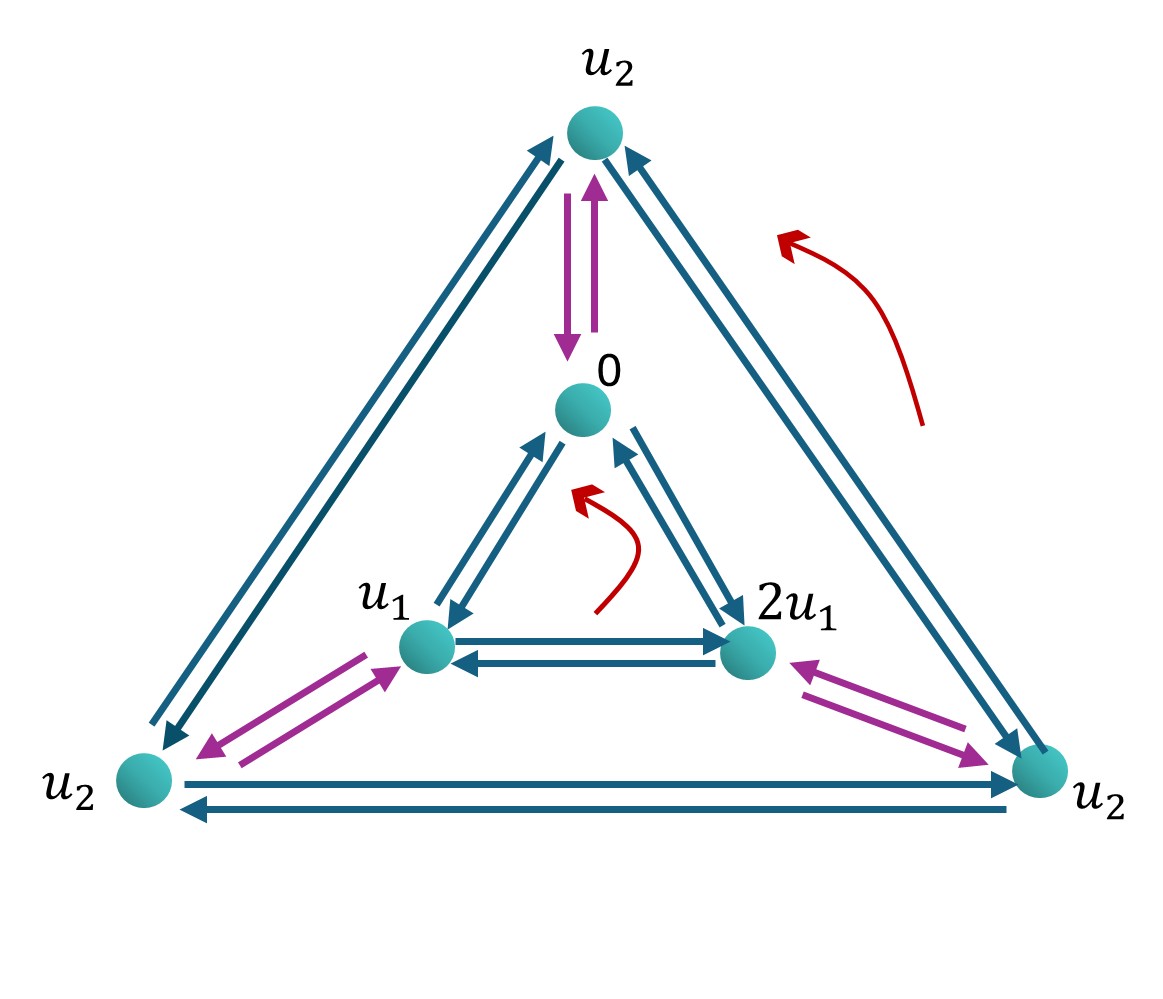}
        \caption{{}}
        \label{dismm3b}
     \end{subfigure}
    \caption{{\bf Sketch of discrete flashing ratchet model for molecular motor motion (a) and its `wrapped' version (b).} \small{ Molecular motor dynamics in discrete space, with the dynamics switching between two potential landscapes. The potential  switches at random times between flat (space independent) and periodic  (space-asymmetric) shapes at a  switching rate $\alpha$ independent of the motor position. The motor is subject to an external constant force  $F_{\text{load}}$ (red arrow) whose strength is independent of the state of the potential. The transitions within each potential follow Eq.~\eqref{eq:bldb}. }
    }
    \label{dismm3}
\end{figure}

To get a more detailed understanding of the molecular motor dynamics and the dependence of its heat-capacity features on the dynamical parameters, we focus now on a  discrete version of the model introduced in~\cite{roldan2010}.
Consider a Markov jump process defined on the graph shown in \fig\ref{dismm3a}. Each state is labeled by a pair $(x,y)$, where $y \in \{1,2\}$. For $y=1$, the system is subject to a skewed periodic potential and for $y=2$ the potential is flat. An external force is acts on the system regardless of the potential state. For each value of $y$, the position $x$ ranges over a set of $3\tau+1$ distinct states, where $\tau$ denotes the periodicity of the potential. The potentials $U(x,1)\in \{0, u_1, 2 u_1\}$ for the states in  $y=1$ are shown in \fig\ref{dismm3a} and for larger values of $\tau$, the potentials are periodically repeated, to be compared with the continuous version of the molecular motor dynamics.

An  equivalent picture that captures the periodicity of this potential consists of two connected cycles, each containing three states, see \fig\ref{dismm3b}. The inner cycle ($y=1$) has potentials $\{0, u_1, 2u_1\}$, while all states in the outer cycle  ($y=1$)  have the same potential $u_2$. All states are subjected to a counterclockwise driving force.  The transition rates for the Markov jump process over each cycle of  the graph in  \fig\ref{dismm3b} are 
\begin{align}\label{eq:bldb}
 k_y(x, x')=\exp\left[\frac{\beta}{2}(U_y(x)-U_y(x')+w(x,x'))\right]. 
\end{align}
 The transition rates for switching the cycles $y=1\to y=2$ and $y=2\to y-1$ are constant $\alpha$ and independent of $x$.
  The function $U_y(x)$ represents the potential of each state, as indicated in Fig.~\ref{dismm3b}. The function $w(x,x')$ gives the work performed by the external force $F_{\text{load}}$ upon the displacement of the motor by one lattice site $x\to x'$. It is given by  $w(x,x')  = w_{\text{load}}$ for jumps in the clockwise direction and $w(x,x') = -w_{\text{load}}$ for jumps in the counterclockwise direction. 

\begin{figure}[H]
	\centering
      \begin{subfigure}{0.49\textwidth}
         \centering
         \def\svgwidth{0.8\linewidth}        
        \includegraphics[scale = 0.85]{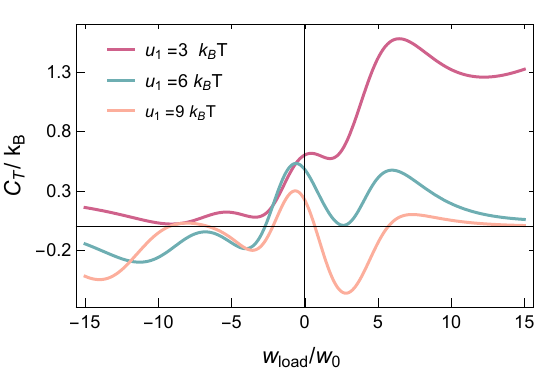}
        \caption{}\label{2hcalphau-a}
     \end{subfigure}
      \hfill
     \centering
      \begin{subfigure}{0.49\textwidth}
         \centering
         \def\svgwidth{0.8\linewidth}        
        \includegraphics[scale = 0.85]{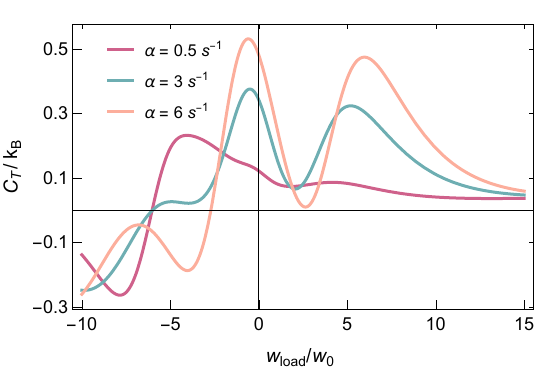}
          \caption{}\label{2hcalphau-b}
     \end{subfigure}
     	\begin{subfigure}{0.49\textwidth}
     	\centering
     	\def\svgwidth{0.8\linewidth}        
     	\includegraphics[scale = 0.85]{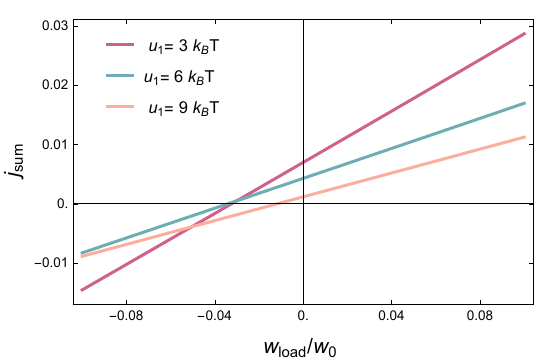}
     	\caption{}\label{2hcalphau-c}
     \end{subfigure}
     \hfill
     \centering
     \begin{subfigure}{0.49\textwidth}
     	\centering
     	\def\svgwidth{0.8\linewidth}        
     	\includegraphics[scale = 0.85]{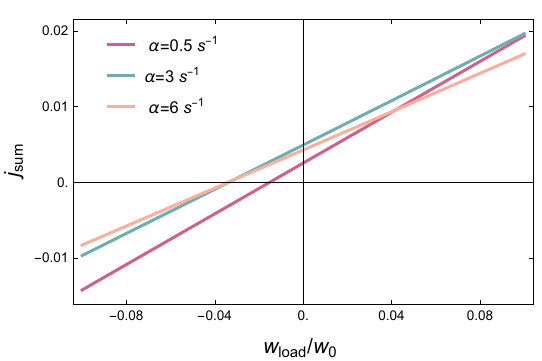}
     	\caption{}\label{2hcalphau-d}
     \end{subfigure}
\caption{\small{{\bf Heat capacity $C_T$ (a,b) and total spatial current  $j_{\text{sum}}$ (c,d) of the discrete flashing ratchet of molecular motor motion as a function of the external load work  $w_{\text{load}}$:}} for different values of the barrier height $u_1$  at fixed $\alpha=6$(a,c) and for different values of the   energy injection rate $\alpha$ at fixed $u_1=6 k_{B}T$(b,d). Here $w_0=k_{B}T$. }\label{2hcalphau}
\end{figure}

As in the diffusion case, even in the absence of \(w_{\text{load}}\) the dynamics can sustain a steady net current $j$ (proportional to the motor's velocity) in the clockwise direction  for $\alpha>0$ as a result of broken detailed balance~\cite{roldan2010}. We report in Fig.~\ref{2hcalphau-c} and Fig.~\ref{2hcalphau-d},  the value of the net current as a function of $w_{\text{load}}$  for different values of $\alpha$ and $u_1$. Here and in the following, we put $j= j_1(x,x+1)+j_2(x,x+1)$  on a specific site $x$ the stationary probability currents from $x\to x+1$ in each cycle ($y=1$ and $y=2$) which yields the total spatial current along $x$.  Extra information and additional results for this model are given in Appendix~\ref{jve}.

\begin{figure}[H]
\centering
      \begin{subfigure}{0.49\textwidth}
         \centering
         \def\svgwidth{\linewidth}        
        \includegraphics[scale = 0.85]{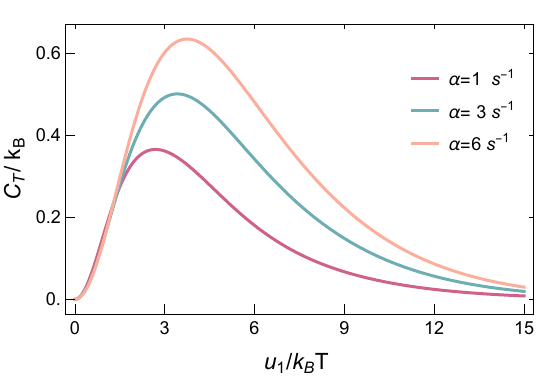}
        \caption{{}}
        \label{Ctvsu1}
     \end{subfigure}
      \hfill
     \centering
      \begin{subfigure}{0.49\textwidth}
         \centering
         \def\svgwidth{\linewidth}        
        \includegraphics[scale = 0.85]{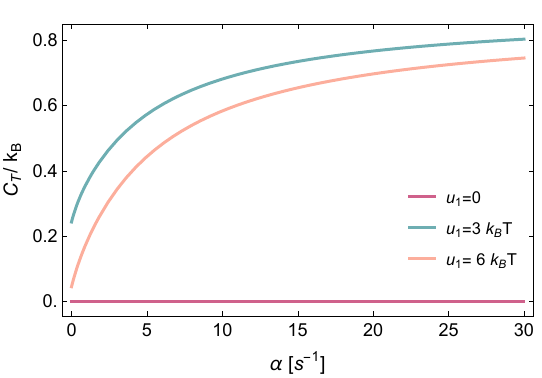}
        \caption{{}}
        \label{Ctvsu2}
     \end{subfigure}
\caption{\small{{\bf Heat capacity of discrete flashing ratchet: spontaneous  $w_{\text{load}}=0$ nonequilibrium fluctuations:} (a)   as a function of the barrier height $u_1$ for different values of the switch rate ($\alpha$, see legend); and (b)  as a function of the switching rate $\alpha$ for different values of the barrier height ($u_1$, see legend),  $\beta=1$.}}   \label{2hcalphau0}
\end{figure}

We first focus in Fig.~\ref{2hcalphau0} on the (external) force-free scenario corresponding to  spontaneous nonequilibrium fluctuations at  $w_{\text{load}}=0$. To this aim, we evaluate the nonequilibrium heat capacity $C_T$ as in Sec.~\ref{dis} for the discrete rower model, i.e., using the quasipotential approach detailed in Appendix~\ref{M1}. First, we analyze how $C_T$ varies with the potential barrier height $u_1$ for different values of the energy injection rate $\alpha$ (Fig.~\ref{Ctvsu1}), finding, as for the discrete rower model,  a non-monotonic relation with  the magnitude  of energy injection   in $k_{\text{B}}T$ units reminiscent of the Schottky anomaly. In addition, the heat capacity's maximal value are, at odds with the discrete rower model (cf. Figs.~\ref{CforTa} and~\ref{CforTb}), increasing with the energy injection rate $\alpha$, and a similar phenomenology of $C_T$ increasing with  $\alpha$ at fixed barrier height (Fig.~\ref{Ctvsu2}).
 An intuitive understanding of the two limiting regimes for $\alpha$ at $w_0=0$ can be obtained by comparing slow and fast switching. In the limit $\alpha \to 0$, the system relaxes within a single landscape before switching occurs. The dynamics therefore approaches equilibrium behavior, and the heat capacity reduces to its equilibrium value, $
C_T^{\text{eq}}=\lim_{\alpha \to 0} C_T(w_0=0).$  In contrast, in the limit $\alpha \to \infty$, the system switches rapidly between the two energy landscapes. Rather than approaching equilibrium, the dynamics experiences an effective averaged topology generated by the fast bidirectional switching. In this regime, $\alpha$ no longer controls relaxation but instead modifies the structure of the state space. The heat capacity approaches a different constant value, $C_T^{0} \equiv \lim_{\alpha \to \infty} C_T(w_0=0),$ which is clearly distinct from the equilibrium result.

The ratio between the nonequilibrium fast-switching $C_T^0$ and equilibrium $C_T^{\text{eq}}$ heat capacities 
increases monotonically with the barrier parameter $\zeta=\beta u_1$ and can be bounded as
\[
\frac{C_T^{0}}{C_T^{\text{eq}}} \ge \frac{1}{2} + \frac{\zeta^2}{4},
\]
with the minimum value $1/2$ attained at vanishing barrier height. This result shows that fast switching generically enhances the heat capacity compared to equilibrium, and that the deviation becomes stronger for increasing barrier strength. Physically, rapid bidirectional switching prevents equilibration within a single landscape and maintains additional fluctuations, which manifest as an increased heat capacity.\\

 Next, we study the heat capacity in the presence (and as a function of) an external load work $w_{\text{load}}$.
In Fig.~\ref{2hcalphau-a} and Fig.~\ref{2hcalphau-b}, the heat capacity as a function of the loaded work is plotted for different values of $u_1$ and $\alpha$. In Fig.~\ref{2hcalphau-c} and Fig.~\ref{2hcalphau-d}, the steady current $j_{\text{sum}}$ is plotted for the same parameter values.  As in the case of the continuous flashing ratchet (see Figs.~\ref{hcr2lm100a} and~\ref{hcr2lm100b}), we find that  
the heat capacity associated with  discrete ratchet model also develops a non-trivial dependency on the external force with multiple peaks, with  local minima ocurring near the stall force. 

\section{Conclusions}
\label{sec:conc}
Biological systems are more than just a material. Our work reinforces the idea that one can use calorimetry to understand, beyond the thermal response by itself, the influence of physiological conditions on biological functioning.  Heat capacity measurements may be an instrument of diagnosis and differentiation. The present paper has applied the main idea of nonequilibrium calorimetry to  relevant models of the physics of life at the microscopic level which can serve as test cases for experimental guidance.\\

We end with some additional remarks. It must be emphasized that implementation requires various experimental challenges to be met.  Having a steady biological system is not as simple as preparing a stationary state  of dead matter.  Time is often limited for observation, and one needs to make sure that the proper physiological conditions can be kept constant while modulating the bath temperature.  Moreover, for micro/meso-biological processes described by one and two degrees of freedom, we have estimated heat capacities of the order of the Boltzmann constant $~k_B$ (i.e., comparable to the heat capacity of {\em a single biomolecule}~\cite{hamzi2022learning}), which follow from analyzing tiny heat fluxes several orders of magnitude below the current detection limits in single-cell microcalorimetry~\cite{fessas2017isothermal,Arunachalam2023, Rodenfels2019, Hong2020}. Related to that and as an outlook, it will  be interesting to continue these studies
to stochastic models of biological processes that result from coarse-graining a large amount of interacting nonequilibrium  variables. Additional computational power may be needed to make contact with the corresponding experiments. \\
 
 {\bf Acknowledgments:}  FK and CM are grateful for the pleasant atmosphere and hospitality at the Quantitative Life Sciences of ICTP, Trieste, where this work was started.  FK is supported by the Research Foundation - Flanders (FWO) postdoctoral fellowship 1232926N. We thank Adrielle Cusi, L\'ea Bresque, and Mohau Pholeli for fruitful discussions.

\appendix
\section*{APPENDIX}

\section{Theory of nonequilibrium heat capacity}
\label{sec:2}
 To make our theoretical analysis more general, both diffusion processes and Markov jump processes are used as modeling tools. Whatever the framework, a central ingredient is the condition of local detailed balance \cite{ldb}, which, for diffusion processes, is verified from the Einstein relation.  Alternatively, we can use the First Law of thermodynamics, which gives the heat and derived concepts such as (expected) heat flux, once we identify the change of energy and the work contribution.
 \subsection{Stochastic models}

\subsubsection{Overdamped diffusion processes}
We consider  systems characterized by two real-valued coordinates $(x_t,y_t)$  undergoing a joint dynamics, with damping $\gamma>0$ and in contact with a thermal bath at temperature $T$ represented here by the standard (zero mean, unit variance) Gaussian white noise $\xi_t$,
\begin{equation}\label{m2}
	\gamma \dot{x}_t = \cal{F}(x_t,y_t;t) - \partial_x U(x_t,y_t) + \sqrt{2\gamma k_{B} T}\, \xi_t,
\end{equation}
where $k_B$ denotes Boltzmann's constant. 
We leave the (supposedly first-order) dynamics of~$y_t$ unspecified so far; in many scenarios it is essential to add feedback so that also its dynamics depends on $x_t$.
In short, we call $x_t$ the system (variable) and think of it as a position, a collective variable, or a summary of the chemomechanical configuration. It is {\em agitated} by its coupling with $y_t$ through a potential $U$ and/or possibly by a time-dependent nonconservative force $\cal F$; for example, an external time-periodic signal that does not even need to depend on $(x,y)$.    Higher-dimensional versions or curved geometries are conceptually similar and are not considered here. Throughout the paper, we use subindex $t$ (e.g. $x_t$)  for stochastic processes evaluated at time $t$, while parenthesis $t$ for quantities that depend deterministically on time $t$. From now on, we write $\partial_x G(x,y) = G'(x,y)$ for the partial derivative with respect to $x$ for functions $G=G(x,y)$.\\
We take $U(x,y)$ to be the system's energy associated with configuration $(x,y)$.  From  It\^o's Lemma we have
\begin{eqnarray}\label{dup}
U(x_t+\id x_t,y_t) - U(x_t,y_t) &=& U'(x_t,y_t)\id x_t + (1/2) U''(x_t,y_t) (\id x_t)^2
\nonumber\\ &=&\mu[({\cal F}(x_t,y_t;t) -  U'(x_t,y_t))U'(x_t,y_t)+ k_B T U''(x_t,y_t)]\id t \nonumber\\ &&+\sqrt{2\gamma k_B T}\,U'(x_t,y_t)\, \xi_t\id t \label{eq:itoU}
\end{eqnarray}
with $\mu=1/\gamma$ the mobility, and  we put $\xi_t\id t =\id B_t$ for the standard Wiener process $B_t$. Therefore, the expected energy change per unit time at time $t$, conditioned on $x_t = x$ and $y_t=y$, reads
\begin{equation}
    \dot{\cal{U}}(x,y;t) = \mu[({\cal F}(x,y;t) -  U'(x,y))U'(x,y)+ k_{B} T U''(x,y)].
\end{equation}
We use analogous definitions for the conditional expectations of work and heat flux rates.\\ 

Secondly, we need the work done by the force ${\cal F}$. The power  $ \dot{\cal{W}}(x,y;t)$ exerted on the system at time $t$ when the state is $(x,y)$ is given by
\begin{align}\label{heateqw}
\dot{\cal{W}}(x,y;t) =& \langle {\cal F} (x,y;t)\circ \dot x\rangle \nonumber\\ 
=&\langle {\cal F} (x,y;t) \dot x\rangle +\mu k_B T {\cal F}\,' (x,y;t) \nonumber \\
=&\mu[ {\cal F}^2 (x,y;t) - {\cal F} (x,y;t)U'(x,y) + k_{B}T {\cal F}'(x,y;t)] .
\end{align}
In the first line we have used the definition of the expected work with $\circ$ denoting the Stratonovich product~\cite{ken}; in the second line we have used the conversion rule between the Stratonovich and the It\^o product; and in the third line we have used Eq.~\eqref{m2} and the rules of It\^o calculus. See  Refs.~\cite{pigolotti2017generic,Roldan2024} for further mathematical details and generalizations to higher dimensions.

Knowing the expected energy change and the expected power, we find the expected heat flux to the bath $ \dot{\cal{Q}}(x,y;t)$ from applying the First Law, $\dot{\cal{Q}}(x,y;t) =\dot{\cal{W}}(x,y;t)-\dot{\cal{U}}(x,y;t) $.  Hence, the expected heat flux  $ \dot{\cal{Q}}(x,y;t)$ at time $t$ from the system to the thermal bath when the state is $(x,y)$ equals
\begin{align}\label{heateq}
\dot{\cal{Q}}(x,y;t) 
 = \mu \Big[  ({\cal F}(x,y;t)-U'(x,y))^2 + k_B T( {\cal F}'(x,y;t) -U''(x,y)) \Big].
\end{align}
This equation may be written in a more compact form as 
\[\dot{\cal{Q}}(x;y;t) = \mu [ F^2_x(x,y;t) + k_{B}T F'_x(x,y;t)]\]
with $F_x(x,y;t)={\cal{F}(x,y;t)-U'(x,y)}$ the total `force' acting on $x$. 

\subsubsection{Markov jump processes}\label{jupr}
Next, consider an irreducible Markov jump process $X_t$ on a discrete state space $K$. System states are denoted by $x, x', \ldots \in K$, and the transition rate for a jump from state $x$ to state $x'$ is $k_y(x, x')$, where $y$ is another dynamical variable (left unspecified for now). Again, 
the jump process may depend on $y$ having their own dynamics with feedback.

The backward generator of such a Markov jump process acting on a function $f = f(x,y)$ reads
\begin{equation}\label{jl}
L_x f(x,y) = \sum_{x'} k_y(x, x')\, [f(x',y) - f(x,y)].
\end{equation}
We need the heat $Q_y(x\to  x')$ released to the thermal bath during a jump from $x$ to $x'$ given $y$, which, from local detailed balance \cite{ldb}  reads 
\begin{equation}\label{heatjump}
Q_y(x \to x') = k_{B}T\log\left(\frac{k_y(x, x')}{k_y(x', x)}\right),
\end{equation}
where here and in the following we use $\log$ for natural logarithm.
The expected heat flux at state $(x,y)$ is
\begin{eqnarray}\label{heateqdis}
    \label{jh}
    {\dot{\cal Q}}(x,y)&=&\sum_{x'} k_y(x,x')\, Q_y(x\to x'),\nonumber\\
    &=& k_{\mathrm{B}}T\sum_{x'} k_y(x,x')\,\log\left(\frac{k_y(x, x')}{k_y(x', x)}\right),
\end{eqnarray}
which is the analogue of Eq.~\eqref{heateq} to discrete states but here restricted to time-independent rates. 

\subsection{Definition of nonequilibrium heat capacity}

Readers may notice that the heat flux ${\dot{\cal Q}}(x,y) = {\dot{\cal Q}}_T(x,y)$ in \eqref{heateq} or  \eqref{heateqdis} depends on the bath's temperature~$T$ which (ignoring other intrinsic time-dependence now for simplicity) may change when we slowly change  $T\rightarrow T+\id T$.  That change is in itself also creating heat, in excess with respect to the stationary power dissipated at the instantaneous temperature.  Its time-integral is called the excess heat during the quasistatic transformation, of the form
\begin{equation}\label{exc}
\delta {\cal Q}^\text{exc}=
\int_0^\tau \id t \,\left[\left\langle {\dot{\cal Q}}_{T(t/\tau)}(x_t,y_t) \right\rangle_t -  \left\langle {\dot{\cal Q}}_{T(t/\tau)}(x,y) \right\rangle \right], \qquad \tau \to \infty. 
\end{equation}
The first expectation $\langle \cdot \rangle_t$ in the integral is for the time-dependent process $(x_t,y_t)$ where the temperature follows a slow protocol  (with $\tau \gg 1$) from $T(0) =T$ to reach a final temperature $T(1) = T+\id T$.  The initial state is drawn from the (unique) stationary distribution at an initial bath temperature $T(0) = T$.  The second expectation in the integral is in the (unique) stationary distribution at (fixed) temperature $T(t/\tau)$.  In that way, the excess heat 
${\cal Q}^\text{exc}$ is a renormalized heat to the thermal bath. The nonequilibrium heat capacity at temperature $T$ is defined (in terms of the heat to the system) as Eq.~\eqref{cdef}, copied here for convenience
\begin{equation}\label{cdef2}
C_T = -\frac{\delta {\cal Q}^\text{exc}}{\id T} .
\end{equation}
  
\subsection{AC-calorimetry}

 We mainly follow \cite{calo}, where AC--calorimetry is introduced for nonequilibrium systems. Fig.~\ref{fig:temp_heat_2x2} and \fig \ref{comp} show the scheme of the AC--calorimetry setup. The driven or agitated sample is in contact with a thermal bath with slowly oscillating temperature, and a thermopile (Peltier element) is measuring the dissipated power.
More precisely, we modulate the bath temperature with (small) frequency $\omega_b\neq 0$, {\it e.g.}, following a harmonic modulation with amplitude $\epsilon_b $ around $T$, that is,
\begin{equation}\label{eq:TperMaes2}
T_b(t)=T[1-\epsilon_b\sin(\omega_b t)].  
\end{equation}
After the system relaxes (in a time that we assume is short compared with the ratio of excess heat to steady power dissipation), we measure the time-dependent heat flux $\dot{\cal Q}(t)$. 
For time-periodic temperatures \eqref{eq:TperMaes2}, in  linear order and neglecting $O(\omega_b^2,\epsilon_b^2)$, we have the relation
\begin{equation}\label{mama2}
   \dot{\cal Q}(t) = \dot{\cal Q}_T +\epsilon_b\, T\,[B_T\, \sin(\omega_b t) + C_T\,\omega_b\, \cos(\omega_b t)],
\end{equation}
for $\delta T=-\epsilon_b T$, and $B_T\delta T = \dot{\cal Q}_T -\dot{\cal Q}_{T+ \delta T} $, while $C_T$ is the nonequilibrium heat capacity as in \eqref{cdef2}; see Fig.~\ref{comp}. We thus obtain the following formula for the nonequilibrium heat capacity in the form of a Fourier coefficient 
\begin{equation}\label{cac}
C_T = \frac 1{ \pi\epsilon_b T}\int_0^{2\pi/\omega_b} \id t\, \cos(\omega_b t) \,[\dot{\cal Q}(t) -\dot{\cal Q}_T],
\end{equation}
which holds for $\omega_b\rightarrow 0$ and $\epsilon_b \rightarrow 0$.  See the methodological points in Appendix \ref{meto}.\\

In the case of a time-periodic driving force like ${\cal F} = {\cal F}_0\sin(\omega_F t)$ in \eqref{m2}, the formula \eqref{cac} is unchanged except that now $\dot{\cal Q}(t)$ not needs to be periodic but mixes the frequencies $\omega_F$ of the driving force and $\omega_b\ll \omega_F$ of the bath, while $\dot{\cal Q}_T = \dot{\cal Q}_T(t)$ becomes periodic with frequency $\omega_b$; see also \cite{elena1,elena2}.

\subsection{Quasipotential} \label{M1}
A (more theoretical or computational) method (applicable for time-homogeneous systems) to find the heat capacity proceeds by finding the so-called quasipotential $V$. $V$ is the unique solution to the Poisson equation
\begin{equation}\label{quasi}
    LV(x,y) + \dot{\cal Q}(x,y) - \dot{\cal Q}_T =0,
\end{equation}
associated with the backward generator $L$ for which the stationary average $\langle V \rangle_T = 0$.  As shown in Refs.~\cite{jir,epl,calo,jchemphys,Pe_2012}, the nonequilibrium heat capacity is given by the averaged derivative of the quasipotential with respect to the temperature:
\begin{equation}\label{hca}
C_T = - \bigg\langle \frac{\partial V}{\partial T}\bigg \rangle_T,
\end{equation}
where here and below  $\langle\;\cdot\; \rangle_T $ denotes the steady average for the (partially unspecified) processes of the previous section.

 For overdamped diffusion processes,  the backward generator $L= L_x+L_y$ ({\it e.g.} of \eqref{m2}):  equals
\[
LV(x,y) = \mu[{\cal F}(x,y) - U'(x,y)]V'(x,y) + k_B TV''(x,y) + L_yV(x,y),
\]
where $L_y$ is  the backward generator for the $y$-process, and the expected heat flux may be retrieved from~\eqref{heateq}.  For Markov jump processes, we use the backward generator \eqref{jl},  that is,
\begin{equation}\label{j22}
L_x V(x,y) = \sum_{x'} k_y(x, x')\, [V(x',y) - V(x,y)],
\end{equation}
and the expected heat fluxes from \eqref{heateqdis}, 
see also \cite{mathnernst, pois} for more details.

\subsection{Methodological points}\label{meto}
We collect some extra information and clarifications concerning the AC-calorimetric method as used in the paper.

In the evaluation of the Fourier coefficient \eqref{cac}, one may consider leaving out the direct power $\dot {\cal Q}_T$ in case it is constant in time.  After all, the integral $\int_0^{2\pi \omega_b} \id t \,\cos (\omega_b t) =0$: for every $\omega_b>0$,
\begin{equation}\label{eqa}
\int_0^{2\pi/\omega_b} \id t\, \cos(\omega_b t) \,[\dot{\cal Q}(t) -\dot{\cal Q}_T] = \int_0^{2\pi/\omega_b} \id t\, \cos(\omega_b t) \,\dot{\cal Q}(t)
\end{equation}
Yet, we need to consider that $\dot{\cal Q}(t)$ contains the same constant constribution.  The situation is as depicted in Fig.~\ref{comp}; there is the constant term $\dot{\cal Q}_T$, and we need to evaluate the out-of-phase component as $\omega_b\downarrow 0$. The numerical work prefers to use the left-hand side in \eqref{eqa}, so that we can subtract the constant term by using the same data.  If we would evaluate the right-hand side of \eqref{eqa}, we encounter possibly an error with respect to the (then single) constant term, of the form 
\begin{equation}\label{err}
\text{ error }\, = \int_0^{2\pi/\omega_b} \id t\, \cos(\omega_b t) \,\xi(t) = O(\omega_b^{-1}),
\end{equation}
where $\xi(t)$ is white noise.  The mean of \eqref{err} is zero right, but the variance is $\int_0^{2\pi/\omega_b} \id t\, \cos^2(\omega_b t) = \pi/\omega_b$ that diverges with $\omega_b\downarrow 0$.\\

Obviously, the evaluation of the integrals proceeds via discretization,
We estimate the heat capacity from a single time series sampled with sampling time $\Delta t$ in the interval $[0,\tau_{\text{exp}}]$ containing $N_t=\tau_{\text{exp}}/\Delta t$ data points assuming that $N_t$ the data spans over $N>1$ cycles of period $\tau=2\pi /\omega_b$. Under such assumptions, one may estimate $C_T$ as 
\begin{equation}
    \hat{C}_T=\left(\frac{2}{\omega_b \epsilon_b T}\right)\frac{1}{N_t}\sum_{i=1}^{N_t} \cos(\omega_b t_i )[\dot{\cal Q}_{T_b(t_i)}(x_i,y_i)-\dot{\cal Q}_T] ,
\end{equation}
where here we have used the notations $t_i = i\Delta t$, $x_i=x(t_i)$ and $y_i=y(t_i)$. 


\section{Temperature dependence of the  heat capacity}\label{sec:moreon}
Traditionally, studying the behavior of heat capacity as a function of temperature has provided valuable insights into phenomena such as the Schottky anomaly and phase transitions. In biological systems, however, large variations in temperature are often unrealistic or physiologically irrelevant. Nevertheless, as long as the system remains within an effective temperature range, it is still useful to theoretically explore how the system responds to larger temperature changes, offering insights into its energetic and structural properties.\\
In this section, we specifically study the heat capacity of the models presented in this paper as a function of temperature.
\subsection{Rowing  Markov-jump  model}
\label{rowmore}
\begin{figure}[H]
 \centering
      \begin{subfigure}{0.49\textwidth}
         \centering
         \def\svgwidth{0.8\linewidth}        
        \includegraphics[scale = 0.38]{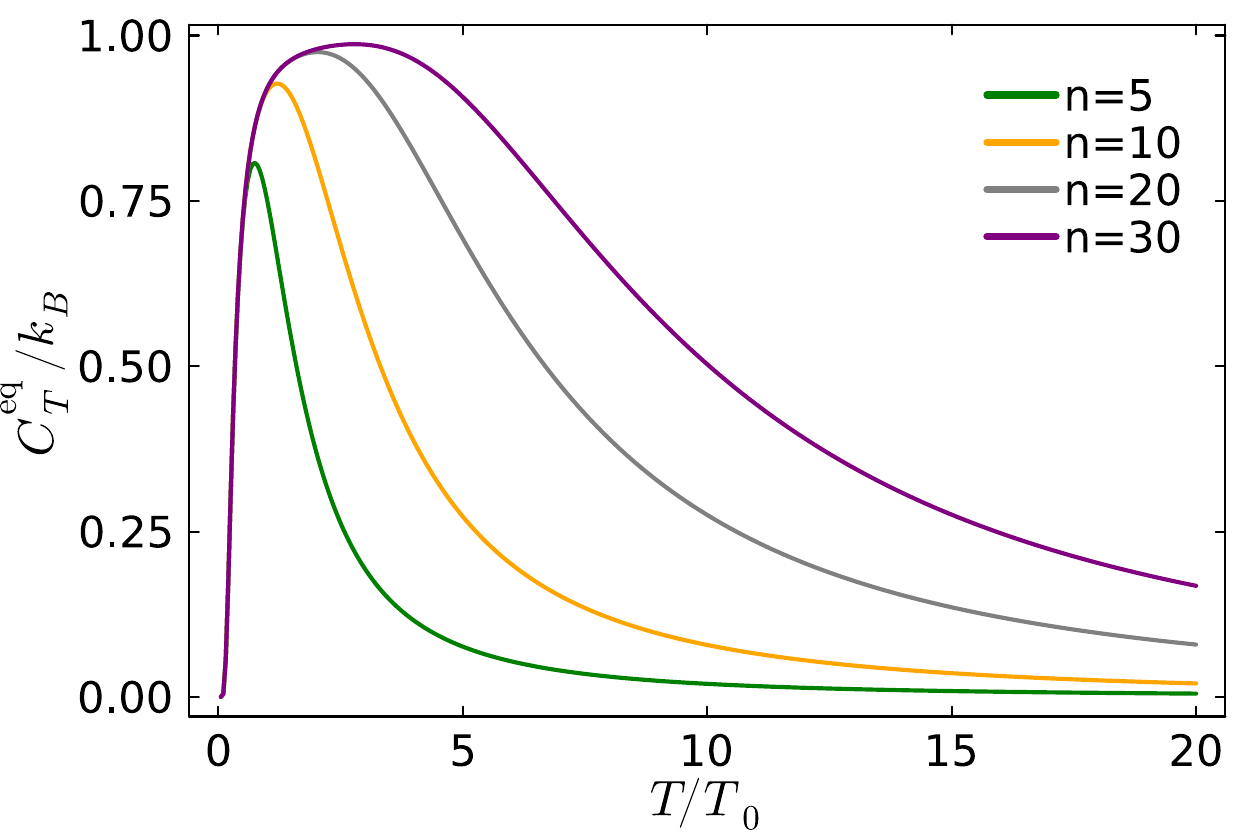}
        \caption{}
     \end{subfigure}
     \hfill
     \centering
      \begin{subfigure}{0.49\textwidth}
         \centering
         \def\svgwidth{0.8\linewidth}        
        \includegraphics[scale = 0.38]{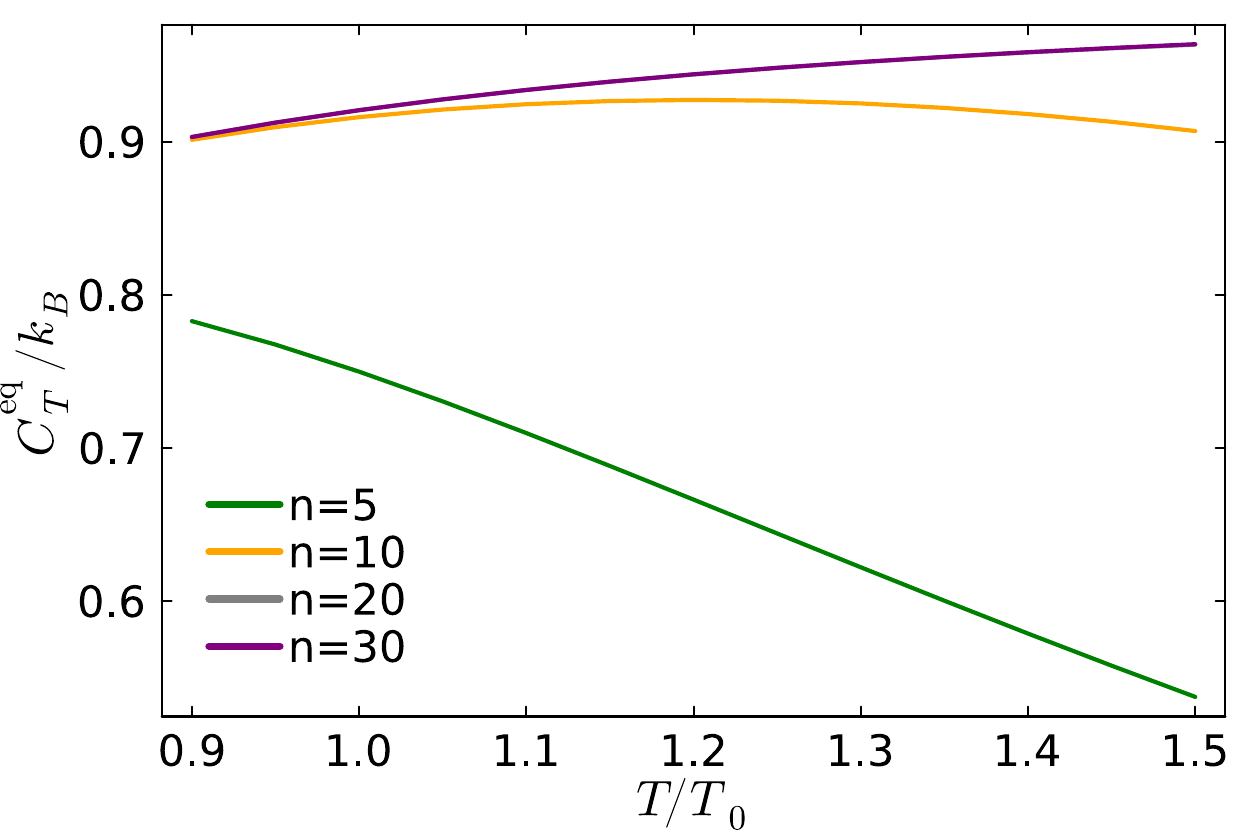}
        \caption{}
     \end{subfigure}
\caption{\small{Nonequilibrium heat capacity for a `dead' rower representing the equilibrium case of the continuous-time Markov jump model sketched in Fig.~\ref{rowjump-a} for different values of $n$, with $\nu =  \nu_0,\, \tau = 1(s) $ and $\alpha = 0$. (a) for a large range of temperatures. (b)  Around room temperatures. For large $n$, lines start to overlap. } }   \label{hceqrowing}
\end{figure}

We provide additional insights to the results in Section \ref{dis}.
In the equilibrium case $\alpha=0$, by a standard calculation, the heat capacity equals
\begin{equation}\label{rowE}
C^{\text{eq}}_T=k_B \, \beta ^2 \nu^2  \left(\frac{ e^{\beta  \nu }}{\left(e^{\beta  \nu }-1\right)^2}-\frac{n^2  e^{n \beta  \nu }}{\left(e^{n \beta  \nu }-1\right)^2}\right),
\end{equation}
and its plot is in Fig.~\ref{hceqrowing}, using the reference $\nu_0=k_B T_0=4 \times 10^{-21} \, J$, where $T_0=290$K denotes  room temperature,  shows the typical Schottky anomaly. 

The peak temperature  $T_\text{peak}$, where the equilibrium heat capacity is maximal, equals
\[
T_\text{peak} = \frac{\nu}{\nu_0} \, y(n)\, T_0,
\] 
where $y(n)$ is a polynomial (almost affine) function of $n$.   As expected for $\nu\simeq \nu_0$, for $n\in [5-15]$, the Schottky peak is around room temperature.

Next we take $\alpha>0$, to compute the nonequilibrium heat capacity using the quasipotential of Section \ref{M1}. That is explicitly possible for any value of $n$. For example, with $n = 3$, the quasipotential takes the values 
\begin{align*}
V(1)&= \frac{1}{Z}\nu  \left(e^{\beta  \nu } \left(e^{\beta  \nu } \left(e^{\beta  \nu } \left(\alpha  \tau  \left(\alpha  \tau -3 e^{\beta  \nu }-5\right)-1\right)-4 \alpha  \tau -3\right)-2 \alpha  \tau -3\right)-2\right),\\
V(2)&= \frac{1}{Z}\nu  \left(e^{\beta  \nu } \left(-e^{\beta  \nu } \left(e^{\beta  \nu }+1\right) \left((\alpha  \tau -1) e^{\beta  \nu }+\alpha  \tau  (\alpha  \tau +2)\right)-2 \alpha  \tau -1\right)-1\right),\\
V(3)&= \frac{1}{Z}\nu  e^{\beta  \nu } \left(e^{\beta  \nu } \left(e^{\beta  \nu } \left((\alpha  \tau +2) e^{\beta  \nu }+\alpha  \tau  (\alpha  \tau +1)+3\right)+2 \alpha  \tau +3\right)+1\right),\\
    Z &=\left(e^{\beta  \nu } \left((2 \alpha  \tau +1) e^{\beta  \nu }+\alpha  \tau +1\right)+1\right)^2.
\end{align*}
The heat capacity is obtained from \eqref{hca}, 
\begin{align*}
C_T&= \frac{k_B}{Z}\beta ^2 \nu ^2 e^{\beta  \nu } \bigg[e^{\beta  \nu } \big(\left(\alpha ^3 \tau ^3+2 \alpha  \tau +5\right) e^{2 \beta  \nu }+(\alpha  \tau  (\alpha  \tau +1)+1) e^{3 \beta  \nu }\\
&\qquad +3 (\alpha  \tau  (\alpha  \tau +1)+2) e^{\beta  \nu }+3 \alpha  \tau +5\big)+1\bigg],\\
    Z &=\left(e^{\beta  \nu } \left((2 \alpha  \tau +1) e^{\beta  \nu }+\alpha  \tau +1\right)+1\right)^3,
\end{align*}
which now also depends on the kinetic parameter $\alpha\,\tau$.\\

\begin{figure}[H]
 \centering
      \begin{subfigure}{0.49\textwidth}
         \centering
         \def\svgwidth{0.8\linewidth}        
        \includegraphics[scale = 0.38]{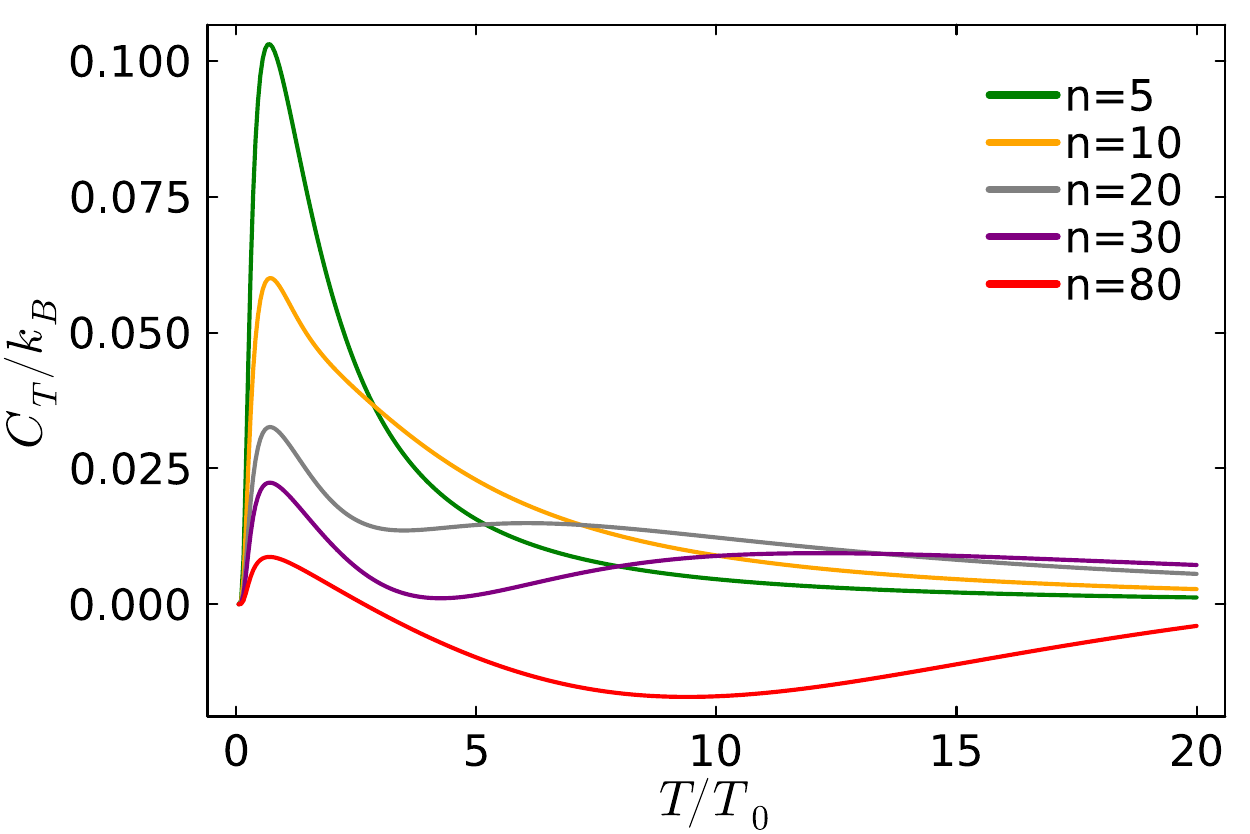}
        \caption{$\alpha =0.5$}
     \end{subfigure}
     \hfill
     \centering
      \begin{subfigure}{0.49\textwidth}
         \centering
         \def\svgwidth{0.8\linewidth}        
        \includegraphics[scale = 0.38]{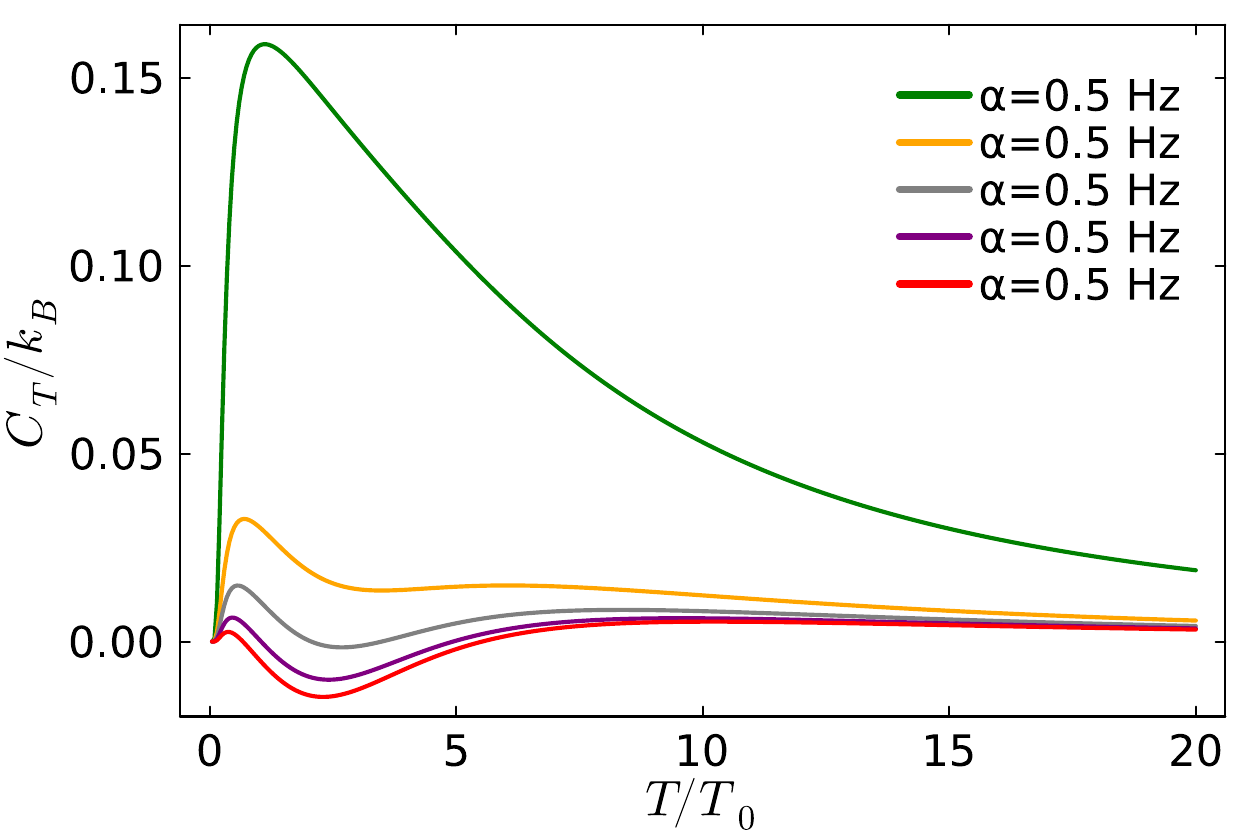}
        \caption{$n=20$}
     \end{subfigure}
\caption{\small{Nonequilibrium  heat capacity of the Markov-jump version of the rowing model, given in~\fig \ref{rowjump-a}, as a function of temperature, for different values of $n$ and $\alpha$, with $\nu =  k_B T_0,\, \tau = 1(s) $, and reference temperature $T_0=290$K. } }   \label{hcrowingdiffn}
\end{figure}

\fig~\ref{hcrowingdiffn} shows  heat capacities for different $ n $ and $\alpha$ varying temperature. As shown in the plots, increasing $\alpha$ and $n$ drives the system further from equilibrium and can lead to negative heat capacity values,  \cite{negheat}.

\subsection{Rowing diffusion model}
The heat capacity as a function of temperature is studied for the diffusion rowing model given in Eq.~\eqref{rowm}.  The plots are comparable to the heat capacity of the jump version, shown in \fig~\ref{hcrowingdiffn}. \\

\begin{figure}[H]
\centering
      \begin{subfigure}{0.49\textwidth}
         \centering
         \def\svgwidth{0.8\linewidth}        
        \includegraphics[scale = 0.9]{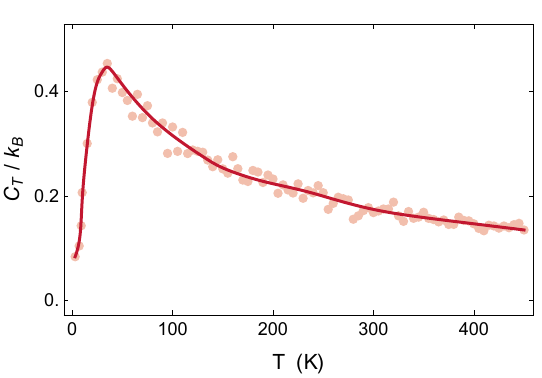}
        \caption{{$\gamma=0.1\,\text{pNs}/\mu \text{m}$}}
     \end{subfigure}
      \hfill
     \centering
      \begin{subfigure}{0.49\textwidth}
         \centering
         \def\svgwidth{0.8\linewidth}        
        \includegraphics[scale = 0.9]{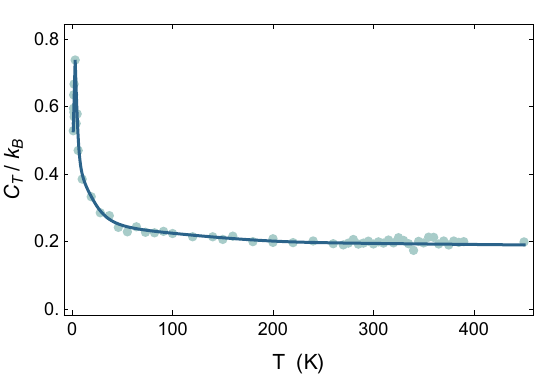}
          \caption{{$\gamma=0.3\,\text{pNs}/\mu \text{m}$}}
     \end{subfigure}
\caption{\small{Nonequilibrium heat capacity of the diffusion rowing model as a function of temperature for differen values of $\gamma$ and $\kappa$. The parameters are  $\kappa=1.5 \,(pN/\mu m),$ $A=1 (\mu m) $, $a=0.25 \, (\mu m)$. The scattered dots represent simulation results obtained from averaging  over $3\times 10^5$ trajectories at each  temperature, while the solid curve is fitted to the data.} }   \label{hcrowing1}
\end{figure}

In Fig.~\ref{hcrowing1}, a wide temperature range from 0 to 450~K is studied, the heat capacity curve exhibits a Schottky peak in both cases, although the peak is located at very low temperatures. For $\gamma = 0.1 \, \mathrm{pN\,s}/\mu\mathrm{m}$, a second peak appears around $T \approx 250 \,\mathrm{K}$. At higher temperatures, the heat capacity increases approximately linearly with temperature. As we show in Figs~\ref{hcrowing1}, the heat capacity  does not vary much with temperature over a range between 300K and 400K.

\begin{figure}[H]
\centering
      \begin{subfigure}{0.49\textwidth}
         \centering
         \def\svgwidth{0.8\linewidth}        
        \includegraphics[scale = 0.85]{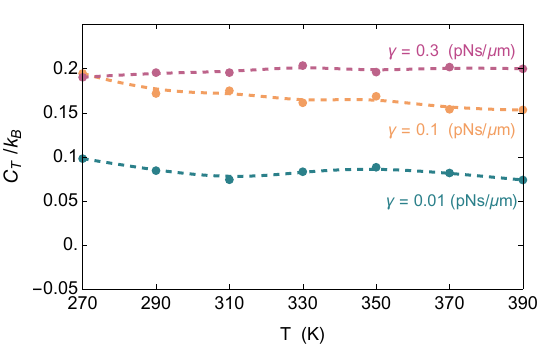}
        \caption{$\kappa=1.5\,\text{pN}/\mu \text{m}$}
     \end{subfigure}
         \begin{subfigure}{0.49\textwidth}
         \centering
         \def\svgwidth{0.8\linewidth}        
        \includegraphics[scale = 0.85]{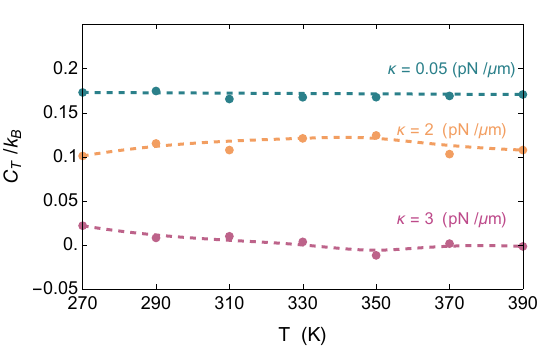}
          \caption{{$\gamma=0.1\,\text{pNs}/\mu \text{m}$}}
     \end{subfigure}
\caption{\small{Nonequilibrium heat capacity of the diffusion rowing model as a function of temperature for different values of $\gamma$ and $\kappa$. The parameters are  $\kappa=1.5 \,(pN/\mu m),$ $A=1 (\mu m) $, $a=0.25 \, (\mu m)$. The scattered dots represent simulation results (averaging  over $3\times 10^5$ trajectories at each  temperature), while the solid curve is fitted to the data.  } }   \label{hcrowing2}
\end{figure}

In Fig.~\ref{hcrowing2}, the heat capacity is plotted for different values of $\kappa$ and $\gamma$ as the temperature varies. The temperature range considered is 270--390~K. As shown in the plots, the heat capacity does not change significantly within this temperature range. It is comparable to the jump version shown in Fig.~\ref{hcrowingdiffn} for the higher temperature range.

\subsection{Molecular motor Markov-jump  model}\label{amm}
\begin{figure}[H]
\centering
      \begin{subfigure}{0.49\textwidth}
         \centering
         \def\svgwidth{0.8\linewidth}        
        \includegraphics[scale = 0.85]{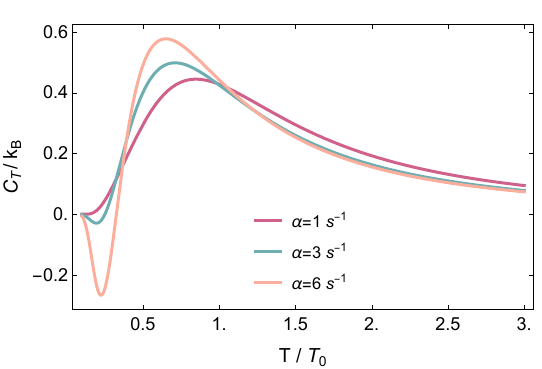}
        \caption{{$w_{\text{load}}= w_0$ }}
     \end{subfigure}
      \hfill
     \centering
      \begin{subfigure}{0.49\textwidth}
         \centering
         \def\svgwidth{0.8\linewidth}        
        \includegraphics[scale = 0.85]{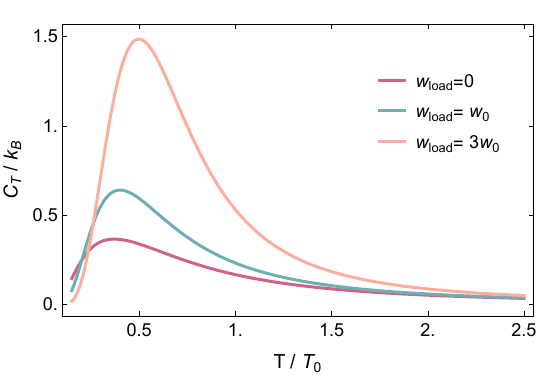}
          \caption{{$\alpha=1$}}
     \end{subfigure}
\caption{\small{The heat capacity of the molecular motor Markov-jump model sketched in \fig\ref{dismm3} as a function of temperature for different values of $\alpha$ and $w_{\text{load}}$ at $u_1=2  k_BT_0$, with $w_0=k_B T_0$, and $T_0=290$K the reference room temperature.} }   \label{2hcmmT}
\end{figure}

The heat capacity of the Markov jump model of the molecular motor as a function of temperature is plotted in Fig.~\ref{2hcmmT}. At lower temperatures, a higher switching rate can lead to a negative heat capacity. A Schottky peak is observed in the heat capacity at intermediate temperatures.

\section{Methodological  point for molecular motor diffusion model} 
\label{app:numericsratchet}
\begin{figure}[H]
\centering
      \begin{subfigure}{0.49\textwidth}
         \centering
         \def\svgwidth{0.8\linewidth}        
        \includegraphics[scale = 0.9]{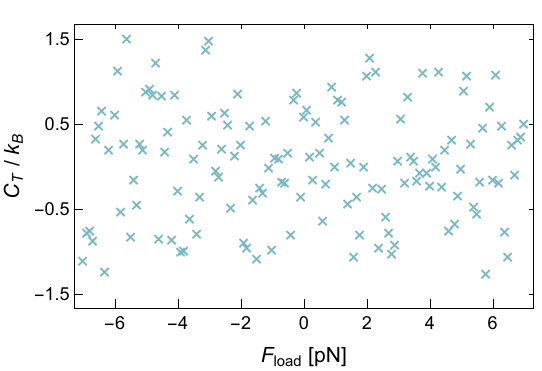}
        \caption{{}} \label{hcr2lm100-a}
     \end{subfigure}
      \hfill
     \centering
      \begin{subfigure}{0.49\textwidth}
         \centering
         \def\svgwidth{0.8\linewidth}        
        \includegraphics[scale = 0.9]{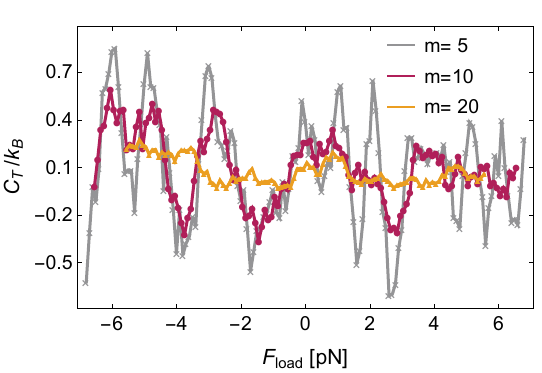}
            \caption{{}} \label{hcr2lm100-b}
     \end{subfigure}
\caption{\small{Heat capacity obtained with AC calorimetry for the molecular motor continuous flashing ratchet model  as a function of the applied load force $F_{\text{load}}$ with parameters given in Table~\ref{tableparam}. (a) Scattered data result from simulations, and (b) a moving average with window sizes $m = 5$, $10$, and $20$ is applied to the data in panel (a).
}}   \label{hcr2lm100}
\end{figure}

As the data obtained from the simulation are very noisy, it is worth mentioning the method used for presenting the data. The simulation is run for molecular motor dynamics to obtain the heat fluxes, taking an average over $10^5$ trajectories during 2000 seconds. The values $\epsilon = 0.02$ and $\omega_B = \pi/5\,\text{Hz}$ at $T = 290\,\text{K}$ are used, while the rest of the parameters are listed in Table \ref{tableparam}, with varying $F_{\text{load}}$. The heat capacity data for different $F_{\text{load}}$ are shown in Fig.~\ref{hcr2lm100-a}. A moving average with window size $m$ is applied to the data in Fig.~\ref{hcr2lm100-a}, and Fig.~\ref{hcr2lm100-b} shows the heat capacity for different values of $m$ with varying  $F_{\text{load}}$. As shown, increasing $m$ reduces  the peak amplitudes.

.

\begin{figure}[H]
\centering
      \begin{subfigure}{0.49\textwidth}
         \centering
         \def\svgwidth{0.8\linewidth}        
        \includegraphics[scale = 0.9]{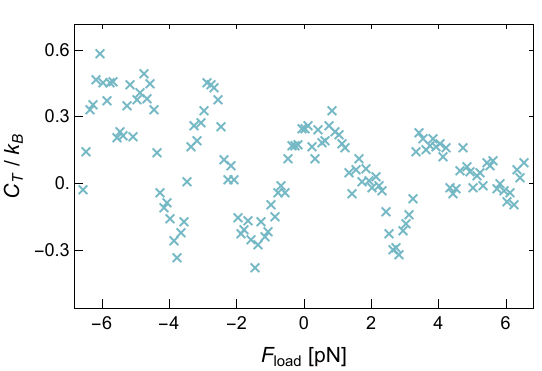}
        \caption{{}} \label{moveandfit-a}
     \end{subfigure}
      \hfill
     \centering
      \begin{subfigure}{0.49\textwidth}
         \centering
         \def\svgwidth{0.8\linewidth}        
        \includegraphics[scale = 0.9]{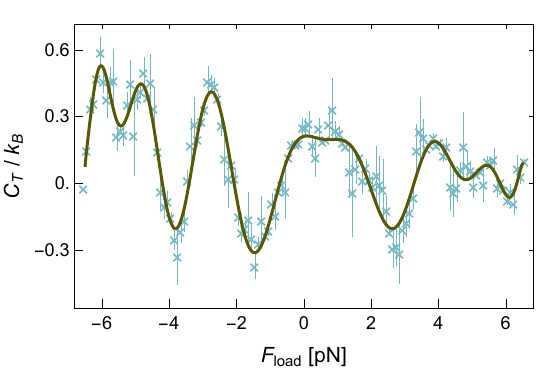}
            \caption{{}} \label{moveandfit-b}
     \end{subfigure}
\caption{\small{Analysis of the data in Fig.~\ref{hcr2lm100-a} after applying a moving average with window size $m=10$ data. 
(a) Heat capacity as a function of the load force after applying a moving average with window size $m=10$ data (purple line in Fig.~\ref{hcr2lm100-b} copied here for convenience. (b) The resulting data are fitted with a polynomial of degree 20, with error bars representing the deviations from the solid curve. The fitted function is $ C_T/k_B = 0.21 + 0.04 F_{\text{load}} + \sum_{i=2}^{20} c_i F_{\text{load}}^i$.}}   \label{moveandfit}
\end{figure}

Fig.~\ref{moveandfit-a} shows the result of applying a moving average with window size $m = 10$ to the data in Fig.~\ref{hcr2lm100-a}. In panel (b), a polynomial of degree 20 is fitted to the data in panel (a).

We notice that the smoothed data exhibit a clear tendency to develop two pronounced peaks, displaying behavior similar to that observed in the discrete version of the model.  In addition, a sudden drop in the  heat capacity is observed at stalling; see Fig.~\ref{hcr2lm100}, where the stalling force is obtained as $F_{\text{stall}} = 1.2~\mathrm{pN}$.


\bibliographystyle{unsrt}  
\bibliography{chr}

@article{pietzonka2016universal,
  title={Universal bound on the efficiency of molecular motors},
  author={Pietzonka, P. and Barato, A. C and Seifert, U. },
  journal={Journal of Statistical Mechanics: Theory and Experiment},
  volume={2016},
  number={12},
  pages={124004},
  year={2016},
  publisher={IOP Publishing}
}

@article{guillet2020extreme,
  title={Extreme-value statistics of stochastic transport processes},
  author={Guillet, A. and Rold{\'a}n, É. and J{\"u}licher, F. },
  journal={New Journal of Physics},
  volume={22},
  number={12},
  pages={123038},
  year={2020},
  publisher={IOP Publishing}
}

@article{baiesi2018life,
  title={Life efficiency does not always increase with the dissipation rate},
  author={Baiesi, M. and Maes, C. },
  journal={Journal of Physics Communications},
  volume={2},
  number={4},
  pages={045017},
  year={2018},
  publisher={IOP Publishing}
}

@incollection{blume1988applications,
  title     = {Applications of Calorimetry to Lipid Model Membranes},
  author    = {A. Blume},
  booktitle = {Physical Properties of Biological Membranes and Their Functional Implications},
  editor    = {Hidalgo, C.},
  series    = {CECs},
  publisher = {Springer},
  address   = {Boston, MA},
  year      = {1988},
  pages     = {71--121},
  doi       = {10.1007/978-1-4613-0935-2_4},
  url       = {https://link.springer.com/chapter/10.1007/978-1-4613-0935-2_4}
}

@article{BLUME1985473,
title = {Calorimetry of lipid model membranes},
journal = {Thermochimica Acta},
volume = {85},
pages = {473-476},
year = {1985},
issn = {0040-6031},
doi = {https://doi.org/10.1016/0040-6031(85)85622-7},
url = {https://www.sciencedirect.com/science/article/pii/0040603185856227},
author = {A. Blume}
}

@article{kolomeisky2007molecular,
  title={Molecular motors: a theorist's perspective},
  author={Kolomeisky, Anatoly B and Fisher, Michael E},
  journal={Annu. Rev. Phys. Chem.},
  volume={58},
  number={1},
  pages={675--695},
  year={2007},
  publisher={Annual Reviews}
}

@article{jeong2001modern,
  title     = {Modern calorimetry: going beyond tradition},
  author    = {Y. H. Jeong},
  journal   = {Thermochimica Acta},
  volume    = {377},
  number    = {1-2},
  pages     = {1--7},
  year      = {2001},
  issn      = {0040-6031},
  doi       = {10.1016/S0040-6031(01)00538-X},
  url       = {https://www.sciencedirect.com/science/article/pii/S004060310100538X},
}

@article{cosentino2001rowers,
  title = {Rowers coupled hydrodynamically. {M}odeling possible mechanisms for the cooperation of cilia},
  author =   {Lagomarsino, M. C. and Bassetti, B. and Jona, P.},
 volume    = {26},
  pages     = {81--88},
  year      = {2002},
  journal= {EPJB},
  year  = {2001},
  url = {https://doi.org/10.1140/epjb/e20020069}
}

@article{chapman1968biological,
  title={Biological membranes},
  author={ D. Chapman},
  year={1974},
  publisher={New York: Academic Press},
  journal = { Thromb Res.},
  volume = {1},
  pages = { 37-40},
  doi = {10.1016/0049-3848(74)90146-7},
}

@article{mathnernst,
      title={The vanishing of excess heat for nonequilibrium processes reaching zero ambient temperature}, 
      author={F. Khodabandehlou and C. Maes and I. Maes and K. Neto\v{c}n\'{y}},
      year={2023},
      url={https://doi.org/10.1007%2Fs00023-023-01367-1},
      doi={10.1007/s00023-023-01367-1},
      issn={1424-0661},
    journal  ={Ann. H. Poincar\'{e}},
}

@article{epl,
doi = {10.1209/0295-5075/96/40001},
url = {https://dx.doi.org/10.1209/0295-5075/96/40001},
year = {2011},
publisher = {},
volume = {96},
number = {4},
pages = {40001},
author = {E.  Boksenbojm and C.  Maes and  K.  Neto\v{c}n\'y and J.  Pe\v{s}ek},
title = {Heat capacity in nonequilibrium steady states},
journal = {Europhys. Lett.},
}

@article{jchemphys,
	doi = {10.1063/5.0142694},
	url = {https://doi.org/10.1063%2F5.0142694},
	year = {2023},
	publisher = {{AIP} Publishing},
	volume = {158}, 
	number = {20}, 
	author = {F. Khodabandehlou and C.  Maes and K. Neto\v{c}n\'{y}},
	title = "{A {N}ernst heat theorem for nonequilibrium jump processes}",
	journal = {J. Chem. Phys.},
}

@article{barato2015thermodynamic,
  title={Thermodynamic uncertainty relation for biomolecular processes},
  author={Barato, A.C and Seifert, U. },
  journal={Physical review letters},
  volume={114},
  number={15},
  pages={158101},
  year={2015},
  publisher={APS}
}

@article{martin2001comparison,
  title={Comparison of a hair bundle's spontaneous oscillations with its response to mechanical stimulation reveals the underlying active process},
  author={P. Martin and AJ. Hudspeth and F. J{\"u}licher},
  journal={Proceedings of the National Academy of Sciences},
  volume={98},
  number={25},
  pages={14380--14385},
  year={2001},
  publisher={The National Academy of Sciences}
}

@article{roldan2024thermodynamic,
  title={Thermodynamic probes of life},
  author={Rold{\'a}n, É.  },
  journal={Science},
  volume={383},
  number={6686},
  pages={952--953},
  year={2024},
  publisher={American Association for the Advancement of Science}
}

@article{manzano2024thermodynamics,
  title={Thermodynamics of computations with absolute irreversibility, unidirectional transitions, and stochastic computation times},
  author={G. Manzano and G.  Karde{\c{s}} and É.   Rold{\'a}n  and D.H. Wolpert},
  journal={Physical Review X},
  volume={14},
  number={2},
  pages={021026},
  year={2024},
  publisher={APS}
}

@article{mizuno2007nonequilibrium,
  title={Nonequilibrium mechanics of active cytoskeletal networks},
  author={D. Mizuno  and C. Tardin and C.F. Schmidt  and F.C. MacKintosh},
  journal={Science},
  volume={315},
  number={5810},
  pages={370--373},
  year={2007},
  publisher={American Association for the Advancement of Science}
}

@article{gnesotto2018broken,
  title={Broken detailed balance and non-equilibrium dynamics in living systems: a review},
  author={F. S Gnesotto  and Mura, F. and Gladrow, J. and Broedersz, C. P},
  journal={Reports on Progress in Physics},
  volume={81},
  number={6},
  pages={066601},
  year={2018},
  publisher={IOP Publishing}
}

@article{wang2015landscape,
  title={Landscape and flux theory of non-equilibrium dynamical systems with application to biology},
  author={Wang, Jin},
  journal={Advances in Physics},
  volume={64},
  number={1},
  pages={1--137},
  year={2015},
  publisher={Taylor \& Francis}
}

@article{loos2023measurement,
  title={Measurement of scale-dependent time-reversal asymmetry in biological systems},
  author={Loos, S. AM},
  journal={Nature Nanotechnology},
  volume={18},
  number={8},
  pages={838--839},
  year={2023},
  publisher={Nature Publishing Group UK London}
}

@article{fernandes2023topologically,
  title={Topologically constrained fluctuations and thermodynamics regulate nonequilibrium response},
  author={Fernandes, M.G. and Horowitz, J.M.},
  journal={Physical Review E},
  volume={108},
  number={4},
  pages={044113},
  year={2023},
  publisher={APS}
}

@article{datta2022second,
  title={Second law for active heat engines},
  author={A. Datta and P. Pietzonka and A.C.  Barato},
  journal={Physical Review X},
  volume={12},
  number={3},
  pages={031034},
  year={2022},
  publisher={APS}
}

@article{horowitz2020thermodynamic,
  title={Thermodynamic uncertainty relations constrain non-equilibrium fluctuations},
  author={Horowitz, J. M.  and Gingrich,T.R. },
  journal={Nature Physics},
  volume={16},
  number={1},
  pages={15--20},
  year={2020},
  publisher={Nature Publishing Group UK London}
}

@article{roldan2021quantifying,
  title={Quantifying entropy production in active fluctuations of the hair-cell bundle from time irreversibility and uncertainty relations},
  author={É.  Rold{\'a}n and J. Barral and P.  Martin  and J. MR Parrondo  and F. J{\"u}licher},
  journal={New Journal of Physics},
  volume={23},
  number={8},
  pages={083013},
  year={2021},
  publisher={IOP Publishing}
}

@article{yang2021physical,
  title={Physical bioenergetics: Energy fluxes, budgets, and constraints in cells},
  author={Yang, X. and Heinemann, M.  and Howard, J. and Huber, G. and Iyer-Biswas, S. and Le Treut, G.  and Lynch, M.  and Montooth, K. L.  and Needleman, D.J  and Pigolotti, S. and others},
  journal={Proceedings of the National Academy of Sciences},
  volume={118},
  number={26},
  pages={e2026786118},
  year={2021},
  publisher={National Academy of Sciences}
}

@Article{ldb,
	title={{Local detailed balance}},
	author={C. Maes},
	journal={SciPost Phys. Lect. Notes},
	pages={32},
	year={2021},
	publisher={SciPost},
	doi={10.21468/SciPostPhysLectNotes.32},
	url={https://scipost.org/10.21468/SciPostPhysLectNotes.32},
}

@Article{activePritha,
	title={{Calorimetry for active systems}},
	author={P. Dolai and C.  Maes and K. Neto\v{c}n\'{y}},
	journal={SciPost Phys.},
	volume={14},
	pages={126},
	year={2023},
	publisher={SciPost},
	doi={10.21468/SciPostPhys.14.5.126},
	url={https://scipost.org/10.21468/SciPostPhys.14.5.126},
}

@article{fodor2016far,
  title={How far from equilibrium is active matter?},
  author={Fodor, {\'E}. and Nardini, C. and Cates, M.E. and Tailleur, J. and Visco, P. and Van Wijland, F.},
  journal={Physical review letters},
  volume={117},
  number={3},
  pages={038103},
  year={2016},
  publisher={APS}
}

@article{oono,
    author = {Y. Oono and M. Paniconi},
    title = "{Steady State Thermodynamics}",
    journal = {Progr. Theor. Phys. Suppl.},
    volume = {130},
    pages = {29-44},
    year = {1998},
    issn = {0375-9687},
    doi = {10.1143/PTPS.130.29},
    url = {https://doi.org/10.1143/PTPS.130.29},
    eprint = {https://academic.oup.com/ptps/article-pdf/doi/10.1143/PTPS.130.29/5213770/130-29.pdf},
}

@article{calo,
	doi = {10.1088/1742-5468/ab4589},
	url = {https://doi.org/10.1088%2F1742-5468%2Fab4589},  
     year = {2019},
	publisher = {{IOP} Publishing},
	volume = {2019},
	number = {11},
	pages = {114004},
	author = {C. Maes and K. Neto{\v{c}}n{\'{y}}},
	title = {Nonequilibrium calorimetry},  
	journal = {J. Stat. Mech.: Theory and Experiment},
}

@article{Pe_2012,
	doi = {10.2478/s11534-012-0053-8}, 
	url = {https://doi.org/10.2478%2Fs11534-012-0053-8},  
	year = {2012}, 
	publisher = {Walter de Gruyter {GmbH}},
	volume = {10}, 
	number = {3},
	author = {J. Pe{\v{s}}ek and E. Boksenbojm and K. Neto{\v{c}}n{\'{y}}},
	title = {Model study on steady heat capacity in driven stochastic systems},
	journal = {Open Physics},
}

@article{pois,
   title={On the {P}oisson equation for nonreversible {M}arkov jump processes},
   volume={65},
   ISSN={1089-7658},
   url={http://dx.doi.org/10.1063/5.0184909},
   DOI={10.1063/5.0184909},
   number={4},
   journal={J. Math. Phys.},
   publisher={AIP Publishing},
   author={F. Khodabandehlou and C. Maes  and K. Neto\v{c}n\'y},
   year={2024} }

@phdthesis{jir,
  author = {J. Pe\v{s}ek},
  title = {Heat Processes in Non-Equilibrium Stochastic Systems},
  year = {2014},
school= {Charles University in Prague}
}

@article{HS,
  author = {T. Hatano and S.-I. Sasa},
  title = {Steady-state thermodynamics of Langevin systems},
  journal = {Phys. Rev. Lett.},
  volume = {86},
  pages = {3463},
  year = {2001},
}

@book{ken,
  author = {K. Sekimoto},
  title = {Stochastic energetics},
  Publisher = {Lecture Notes in Physics},
  volume = {799},
  year = {2010},
}

@article{saitta2022calorimetric,
  title={Calorimetric and thermodynamic analysis of an enantioselective carboxylesterase from Bacillus coagulans: Insights for an industrial scale-up},
  author={F. Saitta and P. Cannazza and S. Donzellaand V. De Vitis and M.  Signorelli and D. Romano and F. Molinari and D. Fessas},
  journal={Thermochimica Acta},
  volume={713},
  pages={179247},
  year={2022},
  publisher={Elsevier}
}

@article{hamzi2022learning,
  title={Learning the hydrophobic, hydrophilic, and aromatic character of amino acids from thermal relaxation and interfacial thermal conductance},
  author={H. Hamzi and A.  Rajabpour and É.   Rold{\'a}n and A.  Hassanali},
  journal={The Journal of Physical Chemistry B},
  volume={126},
  number={3},
  pages={670--678},
  year={2022},
  publisher={ACS Publications}
}

@article{li2020efficiencies,
  title={Efficiencies of molecular motors: a comprehensible overview},
  author={Li, C.B.  and Toyabe, S. },
  journal={Biophysical reviews},
  volume={12},
  number={2},
  pages={419--423},
  year={2020},
  publisher={Springer}
}

@article{Hong2020,
  title = {Sub-nanowatt Microfluidic Single-cell Calorimetry},
  author = {S. Hong and E. Dechaumphai and C. R. Green and R. Lal and A. N. Murphy and C. M. Metallo and R. Chen},
  journal = {Nat. Methods},
  year = {2020},
  volume = {17},
  number = {9},
  pages = {989--993},
  doi = {10.1038/s41592-020-0910-9},
  url = {https://doi.org/10.1038/s41592-020-0910-9},
  publisher = {Nature Publishing Group}  
}

@article{howard2002mechanics,
  title={Mechanics of motor proteins and the cytoskeleton},
  author={Howard, J.  and Clark, R.L.},
  journal={Appl. Mech. Rev.},
  volume={55},
  number={2},
  pages={B39--B39},
  year={2002}
}

@article{keller2000mechanochemistry,
  title={The mechanochemistry of molecular motors},
  author={Keller, David and Bustamante, Carlos},
  journal={Biophysical journal},
  volume={78},
  number={2},
  pages={541--556},
  year={2000},
  publisher={Elsevier}
}

@article{fessas2017isothermal,
  title={Isothermal calorimetry and microbial growth: beyond modeling},
  author={D. Fessas and A. Schiraldi},
  journal={Journal of Thermal Analysis and Calorimetry},
  volume={130},
  number={1},
  pages={567--572},
  year={2017},
  publisher={Springer}
}

@article{Rodenfels2019,
  title = {Heat Oscillations Driven by the Embryonic Cell Cycle Reveal the Energetic Costs of Signaling},
  author = {J. Rodenfels and K. M. Neugebauer and J. Howard},
  journal = {Cell},
  year = {2019},
  volume = {179},
  number = {2},
  pages = {350--364.e15},
  doi = {10.1016/j.cell.2019.08.043},
  url = {https://doi.org/10.1016/j.cell.2019.08.043},
  publisher = {Cell Press}
}

@article{Arunachalam2023,
  title = {Dissecting Flux Balances to Measure Energetic Costs in Cell Biology: Techniques and Challenges},
  author = {E. Arunachalam and W. Ireland and X. Yang and D. Needleman},
  journal = {Annu. Rev. Condens. Matter Phys.},
  year = {2023},
  volume = {14},
  doi = {10.1146/annurev-conmatphys-031620-103841},
  url = {https://doi.org/10.1146/annurev-conmatphys-031620-103841},
  publisher = {Annual Reviews},
}

@article{elena1,
  title = {Heat capacity of periodically driven two-level systems},
  author = {E. Rufeil Fiori and C. Maes},
  journal = {Phys. Rev. E},
  volume = {110},
  issue = {2},
  pages = {024121},
  numpages = {7},
  year = {2024},
  publisher = {American Physical Society},
  doi = {10.1103/PhysRevE.110.024121},
  url = {https://link.aps.org/doi/10.1103/PhysRevE.110.024121}
}

@article{elena2,
  author    = {E. R. Fiori and C. Maes and R. Vidts},
  title     = {Specific Heat of the Driven {C}urie--{W}eiss Model},
  journal   = {J. stat.  Phys.},
  volume    = {192},
  pages     = {63},
  year      = {2025},
  doi       = {10.1007/s10955-025-03438-5},
  url       = {https://doi.org/10.1007/s10955-025-03438-5}
}

@article{pigolotti2017generic,
  title={Generic properties of stochastic entropy production},
  author={S. Pigolotti and I. Neri and É. Rold{\'a}n and F. J{\"u}licher},
  journal={Physical review letters},
  volume={119},
  number={14},
  pages={140604},
  year={2017},
  publisher={APS}
}

@article{Roldan2024,
  title        = {Martingales for Physicists: A Treatise on Stochastic Thermodynamics and Beyond},
  author       = {É. Roldán and I. Neri and R. Chetrite and S. Gupta and S. Pigolotti and F. Jülicher and K. Sekimoto},
  journal      = {Advances in Physics},
  volume       = {72},
  number       = {1-2},
  pages        = {1--258},
  year         = {2024},
  doi          = {10.1080/00018732.2024.2317494},
  url          = {https://doi.org/10.1080/00018732.2024.2317494},
  note         = {arXiv:2210.09983}
}

@article{bruot2016realizing,
  title={Realizing the physics of motile cilia synchronization with driven colloids},
  author={N. Bruot  and P. Cicuta},
  journal={Annual Review of Condensed Matter Physics},
  volume={7},
  number={1},
  pages={323--348},
  year={2016},
  publisher={Annual Reviews}
}

@article{kotar2010hydrodynamic,
  title={Hydrodynamic synchronization of colloidal oscillators},
  author={J. Kotar and M.  Leoni and B. Bassetti and M. Lagomarsino and P. Cicuta},
  journal={Proceedings of the National Academy of Sciences},
  volume={107},
  number={17},
  pages={7669--7673},
  year={2010},
  publisher={National Academy of Sciences}
}

@article{gupta2025,
      title={Role of activity and dissipation in achieving precise beating in cilia: Insights from the rower model}, 
      author={S. Gupta and D.  Chaudhuri and S.  Dey},
      year={2025},
      volume={2504.07681},
       journal={arXiv},
      primaryClass={cond-mat.soft},
      url={https://arxiv.org/abs/2504.07681}, 
}

@article{negheat,
   title={Negative specific heats: where {C}lausius and {B}oltzmann entropies separate},
   volume={27},
   ISSN={1463-9084},
   url={http://dx.doi.org/10.1039/d5cp01269d},
   DOI={10.1039/d5cp01269d},
   number={28},
   journal={PCCP},
   publisher={Royal Society of Chemistry (RSC)},
   author={L. Bogers and F.  Khodabandehlou and C. Maes},
   year={2025},
   pages={15009–15023} }

@article{Ajdari,
  title = {Modeling molecular motors},
  author = {F. J\"ulicher and A.  Ajdari and J. Prost},
  journal = {Rev. Mod. Phys.},
  volume = {69},
  issue = {4},
  pages = {1269--1282},
  numpages = {0},
  year = {1997},
  publisher = {American Physical Society},
  doi = {10.1103/RevModPhys.69.1269},
  url = {https://link.aps.org/doi/10.1103/RevModPhys.69.1269}
}

@article{Fulga2009,
  author    = {F. Fulga and D. V.  Nicolau Jr and D. V. Nicolau},
  title     = {Models of protein linear molecular motors for dynamic nanodevices},
  journal   = {Integr. Biol.},
  volume    = {1},
  number    = {2},
  pages     = {150--169},
  year      = {2009},
  doi       = {10.1039/b814985b},
  pmid      = {20023800},
  publisher = {Royal Society of Chemistry}
}

@article{Astumian2010,
  author    = {Astumian, R. Dean},
  title     = {Thermodynamics and kinetics of molecular motors},
  journal   = {Biophys. J.},
  year      = {2010},
  volume    = {98},
  number    = {11},
  pages     = {2401--2409},
  doi       = {10.1016/j.bpj.2010.02.040},
  pmid      = {20513383},
  pmcid     = {PMC2877326},
}

@article{Reimann2002,
   title={Brownian motors: noisy transport far from equilibrium},
   volume={361},
   ISSN={0370-1573},
   url={http://dx.doi.org/10.1016/S0370-1573(01)00081-3},
   DOI={10.1016/s0370-1573(01)00081-3},
   number={2–4},
   journal={Phys. Rep.},
   publisher={Elsevier BV},
   author={P. Reimann},
   year={2002},
    pages={57–265} }

@article{EnW,
title = {Confusion and clarification: Albert {E}instein and {W}alther {N}ernst's {H}eat {T}heorem, 1911–1916},
journal = {Stud. Hist. Philos. Sci. B: Modern Physics},
volume = {37},
number = {1},
pages = {101-114},
year = {2006},
note = {2005: The Centenary of Einstein's Annus Mirabilis},
issn = {1355-2198},
doi = {https://doi.org/10.1016/j.shpsb.2005.10.001},
url = {https://www.sciencedirect.com/science/article/pii/S1355219805000687},
author = {A. J.  Kox}
}

@inbook{J_licher,
   title={Force and motion generation of molecular motors: A generic description},
   ISBN={9783540480709},
   url={http://dx.doi.org/10.1007/BFb0104221},
   DOI={10.1007/bfb0104221},
   booktitle={Transport and Structure},
   publisher={Springer Berlin Heidelberg},
   author={F. Jülicher},
   pages={46–74},
   year={2007} }

@article{Roldan2010,
   title={Estimating Dissipation from Single Stationary Trajectories},
   volume={105},
   ISSN={1079-7114},
   url={http://dx.doi.org/10.1103/PhysRevLett.105.150607},
   DOI={10.1103/physrevlett.105.150607},
   number={15},
   journal={Physical Review Letters},
   publisher={American Physical Society (APS)},
   author={ É. Roldán and  J.  M. R. Parrondo},
   year={2010},
   month=oct }
\onecolumngrid

\end{document}